\documentclass[a4paper,11pt]{article}
\pdfoutput=1 % if your are submitting a pdflatex (i.e. if you have
\usepackage{jcappub} % for details on the use of the package, please
\usepackage{enumerate}
\usepackage{graphicx}
\usepackage[utf8]{inputenc}

\usepackage[dvipsnames]{xcolor}

\title{Solar Atmospheric Neutrinos and the Sensitivity Floor for Solar Dark Matter Annihilation Searches}

\author[a]{C.A. Arg\"{u}elles}
\author[b]{G. de Wasseige}
\author[c]{A. Fedynitch}
\author[d]{B.J.P. Jones}

\affiliation[a]{Massachusetts Institute of Technology, 77 Massachusetts Ave., Cambridge MA, USA}
\affiliation[b]{Vrije Universiteit Brussel, Pleinlaan 2, 1050 Elsene, Brussels, Belgium}
\affiliation[c]{Karlsruhe Institute of Technology, 76021 Karlsruhe, Germany}
\affiliation[d]{University of Texas at Arlington, 108 Science Hall, 502 Yates St, Arlington TX, USA}

\emailAdd{caad@mit.edu}
\emailAdd{gdewasse@vub.ac.be}
\emailAdd{anatoli.fedynitch@desy.de}
\emailAdd{ben.jones@uta.edu}

\footnotetext[0]{The fluxes of high energy neutrinos from the solar atmosphere using the FJAWs model can be downloaded from \path{http://www-hep.uta.edu/~bjones/FJAWs/index.html} and \path{http://hdl.handle.net/1721.1/108394}}

\abstract{Cosmic rays interacting in the solar atmosphere produce showers that result in a flux of high-energy neutrinos from the Sun.  These form an irreducible background to indirect solar WIMP self-annihilation searches, which look for heavy dark matter particles annihilating into final states containing neutrinos in the Solar core.  This background will eventually create a sensitivity floor for indirect WIMP self-annihilation searches analogous to that imposed by low-energy solar neutrino interactions for direct dark matter detection experiments. We present a new calculation of the flux of solar atmospheric neutrinos with a detailed treatment of systematic uncertainties inherent in solar atmospheric shower evolution, and we use this to derive the sensitivity floor for indirect solar WIMP annihilation analyses. We find that the floor lies less than one order of magnitude beyond the present experimental limits on spin-dependent WIMP-proton cross sections for some mass points, and that the high-energy solar atmospheric neutrino flux may be observable with running and future neutrino telescopes.}

\begin{document}
\maketitle
\flushbottom

\section{Introduction\label{sec:Motivation}: Direct and indirect dark matter searches}
The existence of gravitationally interacting dark matter has been
well established through a wide variety of cosmological and astrophysical
observations \cite{Bertone:1235368}. However, the experimental observation of particle
dark matter remains elusive \cite{Agashe:2014kda}. Among techniques used to search
for particle dark matter are direct detection \cite{Gaitskell:2004gd}, accelerator
production \cite{Feng:2010gw} and indirect astrophysical searches \cite{Feng:2000zu,Arguelles:2017atb}. These techniques
are highly complementary, allowing access to different regions of
dark matter model space.

Direct detection is based on the principle that, if dark matter particles
have interactions beyond gravitational ones with standard model
particles, low-background particle detectors may observe the recoils
of nuclei scattering from dark matter particles in the galactic
halo. The standard paradigm is to assume that dark matter particles
are WIMPs (weakly interacting massive particles), and following assumptions about the velocity and
density distribution of the Galactic halo of dark matter \cite{Green:2011bv}, non-observation of nuclear
recoils in such detectors have allowed limits to be placed on the
WIMP-nucleus cross section. These limits presently extend to $\sigma=10^{-45}\mathrm{cm^{2}}$
at the most sensitive mass of $40\,\mathrm{GeV}$ \cite{Akerib:2016vxi}.
The limits are considerably weaker if the dark matter-nucleus interaction
is spin dependent, extending only to $5\times10^{-40}\mathrm{cm^{2}}$ in a similar mass range \cite{Amole:2016pye, Amole:2015pla}. 

The precision
of many direct dark matter detection experiments is now nearing sensitivity to coherent scatters of low-energy solar neutrinos.  These neutrinos represent a source of background that is effectively irreducible in traditional detectors. The WIMP scattering cross section at which the dark matter signal would be buried under coherent neutrino scattering events is commonly referred to as a ``sensitivity floor'' for direct dark matter detectors. Although a WIMP signal could still be observed over well-predicted background models, working in the large-background regime implies a significant impediment to the rate of experimental progress \cite{Billard:2013qya}, which can make sensitivity improvements in traditional dark matter searches prohibitively slow.  On the other hand, it may be still possible to exploit additional observables, including directional detection \cite{Ahlen:2009ev} and annual
modulation \cite{Freese:2012xd}, to move beyond the limitations imposed by the coherent neutrino background, if suitable detectors can be devised.

Indirect detection is an independent and complementary method of searching for dark matter.
The technique relies on searching for annihilation products of dark
matter particles in astrophysical environments \cite{Silk:1985ax,Press:1985ug}, where the local dark matter density is
expected to be high enough to facilitate copious WIMP self-annihilation. Potentially detectable
annihilation products include neutrinos, gamma rays, positrons,
antiprotons and anti-nuclei \cite{Bertone:1235368}. A channel that has drawn particular attention is the production of
high-energy neutrinos from WIMP self-annihilation in the solar core. The
Sun would act as a concentrator of dark matter particles if and only if WIMPs
interact sufficiently often with regular matter to effectively transfer
their kinetic energy to solar material as they traverse the Sun, causing them to pool near the center. Thus
the strength of this signature dependends on both the self-annihilation
cross section (calculable in WIMP models) and the WIMP-proton cross section. 

Large neutrino detectors, including SuperKamiokande \cite{Aartsen:2012kia,Desai:2004pq}, ANTARES \cite{Adrian-Martinez:2016gti}, and IceCube \cite{Aartsen:2016exj}, have
searched for fluxes of high-energy neutrinos produced through solar WIMP self-annihilation
and set limits on both spin-dependent and spin-independent WIMP-proton couplings.
The limits assume annihilation to an intermediate particle that decays
into neutrinos, with the strength of the limit depending on the final state and its kinematics. Three channels of particular interest are $\chi\chi\rightarrow b\bar{b}$, $\chi\chi\rightarrow W^{+}W^{-}$, and $\chi\chi\rightarrow \tau^{+}\tau^{-}$. In all cases, the limits on the spin-dependent cross section exceed the strength of those from direct searches. At the time of writing, IceCube holds the strongest published spin-dependent cross section limit, using 3 years of data from the full 86-string detector to set limits that peak at $3\times10^{-39}\,\mathrm{cm^{2}}$ for $\chi\chi\rightarrow b\bar{b}$,
 $5\times10^{-41}\,\mathrm{cm^{2}}$ for $\chi\chi\rightarrow W^{+}W^{-}$, and $2\times10^{-41}\,\mathrm{cm^{2}}$ for $\chi\chi\rightarrow \tau^{+}\tau^{-}$  modes \cite{Aartsen:2016zhm}.
IceCube and ANTARES will continue collecting data and improving their limits, and future neutrino telescopes, including PINGU \cite{Aartsen:2014oha} and KM3NeT \cite{Adrian-Martinez:2016fdl}, will also be sensitive to dark matter annihilation signatures.

The experimental technique involves treating the Sun as a point source
and searching for an unexpected flux of high-energy neutrinos from
that direction--we discuss this signature in more detail in Section \ref{sec:Signature}. The backgrounds in the leading searches today are dominated by the
flux of atmospheric neutrinos produced by cosmic ray interactions
in the Earths atmosphere. This background is in principle
reducible by improving the angular resolution of large neutrino detectors, either by improved analysis methods in existing experiments or by design improvements for future detectors.

A second source of background, presently sub-dominant but effectively irreducible,
is the flux of high-energy neutrinos produced by cosmic ray induced showers
in the solar atmosphere.  These neutrinos have been discussed before, for example in \cite{Seckel:1991ffa,Moskalenko:1993ke,Ingelman:1996mj,PhysRevD.74.093004, Hettlage:1999zr}. In this paper we present a new, systematically
detailed calculation of this flux\footnote{We will refer to our calculation as the Fedynitch-Jones-Arg\"{u}elles-Wasseige solar flux, or \emph{FJAWs}}, and we discuss its 
implication as a background to indirect WIMP annihilation searches.  We predict the event rates in existing and proposed neutrino telescopes, and calculate the effective ``sensitivity floor'' imposed by solar atmospheric neutrinos on indirect solar WIMP annihilation searches.

This paper is organized as follows. Section \ref{sec:Signature}
discusses the experimental signature of WIMP self-annihilation and the
method of searching for it with neutrino telescopes. Section \ref{sec:FluxCalc}
details our calculation of the solar atmospheric neutrino flux, from
primary modeling, through shower simulation, oscillation and propagation
effects. In Section \ref{sec:Signal} we predict the event rates in various running and proposed neutrino telescopes, and in Section \ref{sec:Background} we consider the effects of this flux as
a background to WIMP searches, deriving the sensitivity floor in terms of spin-dependent cross section for the $\chi\chi\rightarrow b\bar{b}$, $\chi\chi\rightarrow W^{+}W^{-}$, and $\chi\chi\rightarrow \tau^{+}\tau^{-}$  self-annihilation channels. Finally, in Section \ref{sec:Conclusion} we present our conclusions.

\section{The indirect dark matter signature \label{sec:Signature}}

\subsection{Dark matter annihilation channels}

In order to compare the high energy solar atmospheric neutrino flux to the proposed signals from dark matter self-annihilation, we must specify the dark matter self-annihilation rate and neutrino yield. With these elements, the neutrino spectrum is given by
\begin{equation}
\frac{dN_\nu}{dE_\nu} = \frac{\Gamma^{ann}_\odot}{4\pi d^2} \sum \frac{dN_f}{dE_\nu},
\label{eq:DManna}
\end{equation}
where $d$ is the mean Sun-Earth distance and the neutrino spectrum, $\frac{dN_f}{dE_\nu}$, for dark matter annihilation after solar propagation is taken from \cite{1475-7516-2008-01-021}. The sum is over the dark matter annihilation channels $f$. The annihilation rate is related to the dark matter capture rate by \cite{Cirelli:2005gh}
\begin{equation}
\Gamma^{ann}_\odot = \frac{\Gamma^{capt}_\odot}{2} \tanh^2(t_0/\tau_A)
\label{eq:DMequi}
\end{equation}
where $t_0 = 4.5~ {\rm Gyr}$ is the Sun age and $\tau_A$ is the capture and annihilation time-scale. In our calculation, which is analogous to the procedure described in \cite{Arguelles:2012cf}, we will assume thermal equilibrium so $\Gamma^{ann}_\odot \approx \frac{\Gamma^{capt}_\odot}{2}$. The dark matter capture rate, $\Gamma^{capt}_\odot \propto \sigma_{\nu p}\rho_{\rm DM}$, as a function of the dark matter mass and the dark matter proton scattering cross section, is calculated using formulae from \cite{Gould:1991hx,Jungman:1995df} assuming a spin dependent interaction. We assume a local dark matter density of 0.3 ${\rm GeV}/{\rm cm^3}$ with Maxwell-Boltzmann distribution with a Sun local velocity of 270 km/s \cite{Kerr:1986hz}. Thus, following~\cite{Jungman:1995df}, the capture rate is 
\begin{equation}
\Gamma^{capt} = 1.3\times 10^{25} {\rm s}^{-1} \left(\frac{\rho_{\chi}}{0.3~{\rm GeV} {\rm cm^{-3}}}\right)
\left(\frac{270 {\rm km/s}}{v_{sun}}\right) \left(\frac{\rm GeV}{m_\chi}\right) \left(\frac{\sigma_{\chi p}}{10^{40} {\rm cm}^2}\right) S(\frac{m_\chi}{m_H}),
\end{equation}
where $m_H$ is the hydrogen atom mass and $S$ is kinematic supression factor given in~\cite{PhysRevD.44.3021}.

In this work we consider three different independent annihilation scenarios: ${\rm DM}-{\rm DM} \to b\bar b$, ${\rm DM}-{\rm DM} \to W^+W^-$, and ${\rm DM}-{\rm DM} \to \tau^+\tau^-$. In the $\tau^+\tau^-$ and $W^+W^-$ scenarios, neutrinos are produced promptly by decay, thus producing a hard spectrum. On the other hand, in the $b\bar b$ channel the fermion pair hadronizes into $B$ mesons, which at the high solar densities interact before decaying, thus producing a softer spectrum~\cite{Cirelli:2005gh,1475-7516-2008-01-021}.

\subsection{The WIMP annihilation signature in neutrino detectors}
Neutrinos are ideal candidates to search for WIMP annihilations in celestial bodies such as the Sun or the Earth as their small cross section allows them to escape high density medium without significantly interacting. Since the interaction length becomes smaller than the solar radius for neutrino energies above 1~TeV, the solar WIMP searches are typically performed in an energy range from a few GeV up to a few TeV. 

The event selection criteria are typically optimized to extract charged-current (CC) muon (anti-)~neutrino interactions.  This is because alternative channels, including neutral current and CC electron/tau neutrino interactions suffer from lower angular resolution. Muons from CC muon neutrino interactions produce track-like events, which in this energy range are effectively collinear with the initial neutrino direction. 
It is therefore possible to pinpoint the direction of the Sun by reconstructing the muon track direction, thus allowing the rejection of the terrestrial atmospheric neutrino background.

A night-search, i.e. when the Sun is below the horizon, comes with an advantage of further reducing this background as the Earth would play the role of a shield for atmospheric muons. Day-searches can also be performed by the use of harder cuts in the event selection to limit the atmospheric muon contamination, leading inevitably to a smaller neutrino acceptance compared to night-searches.
Terrestrial atmospheric neutrinos reach the detector in both day and night searches, being rejected using the angular reconstruction described above.

Statistical methods (e.g. likelihood or counting experiment) can then be performed to highlight an excess of neutrino events in the direction of the Sun compared to the rest of the observable sky. A significant excess of events in the angular distribution, sometimes coupled with an energy distribution, would therefore highlight a potential dark matter signal.

To date, no significant excess of events with respect to the expected background has been found. Limits on the solar neutrino flux can therefore be set. Several assumptions are needed in order to express the results in a limit on the WIMP-proton scattering cross section, comparable with direct search results.
Among these are the local dark matter density, the velocity distribution of the halo, and the assumption of equilibrium between capture and annihilation. These determine the density of WIMPs acreted in the solar core.

\section{The solar atmospheric neutrino flux \label{sec:FluxCalc}}

Just as cosmic rays impinging on the Earth's atmosphere produce air showers, leading to hadrons which decay to atmospheric neutrinos~\cite{Gaisser:2002jj}, cosmic rays interacting in the solar atmosphere also produce a high-energy neutrino flux.  This flux was studied in references~\cite{Ingelman:1996mj,PhysRevD.74.093004, Hettlage:1999zr}. The main conclusion was that this flux of neutrinos is small, and therefore unlikely to be useful for detailed study of (for example) neutrino oscillations. However, as the precision of neutrino telescopes improves and increasingly high statistics samples are collected, this flux will assume a new importance as the limiting background to indirect solar WIMP annihilation searches.  In this section we present a new calculation of the solar atmospheric flux, incorporating detailed treatments of systematic uncertainties inherent in neutrino production in the solar atmosphere.

The production of neutrinos in solar showers is different from their production in terrestrial air showers in a few key ways. First, the region of solar atmosphere where the majority of production is localized is significantly less dense and further extended than its terrestrial counterpart. This allows for longer decay lengths of high-energy hadrons before they are absorbed through inelastic interactions, reducing the suppression of the high-energy neutrino flux observed in the Earths atmosphere. On the other hand, the solar core is very large and dense relative to the Earth, so more high-energy neutrinos are lost through interactions when propagating across it. Finally, the path lengths in the solar atmosphere are long enough (thousands of kilometers) that high-energy muons decay and produce a sizeable contribution to the solar atmospheric neutrino flux, whereas in terrestrial air showers these would be stopped abruptly in the Earths crust.

To calculate the solar atmospheric neutrino flux, a model of the solar atmospheric density profile is required. This is discussed in Section~\ref{sec:SolarAtmopshere}. 
The calculation of the solar atmospheric neutrino flux within this density profile is best performed by solving a system of cascade equations~\cite{gaisser1990cosmic}. For this purpose we use the general purpose {\tt MCEq} hadronic cascade solver~\cite{Fedynitch:2015zbe,Fedynitch:2015zma}, which  has previously been used for detailed studies of neutrino production in terrestrial air showers.  We derive not only a prediction of the solar atmospheric flux production in the solar atmosphere but also a suite of systematic uncertainties on this flux in Section~\ref{sec:MCEq}. The decay of muons which punch through the end of the {\tt MCEq} geometry is then simulated using a numerical method, yielding a neutrino flux to be propagated across the higher density solar matter.

Having produced the shower flux, oscillation and propagation effects for neutrinos travelling across the Sun  must be treated. The oscillation calculation accounts fully for the details of the solar matter potential and its coherent and incoherent effects on neutrino oscillations, using the nuSQuIDS software framework~\cite{Delgado:2014kpa}, discussed in Section~\ref{sec:nuSQUIDS}.  After propagation to the Earth, our predicted flux at Earth is presented in Section~\ref{sec:AtEarth}.

\subsection{Modelling of the solar atmosphere \label{sec:SolarAtmopshere}}

\begin{figure}[t]
\begin{centering}
\includegraphics[width=0.9\textwidth]{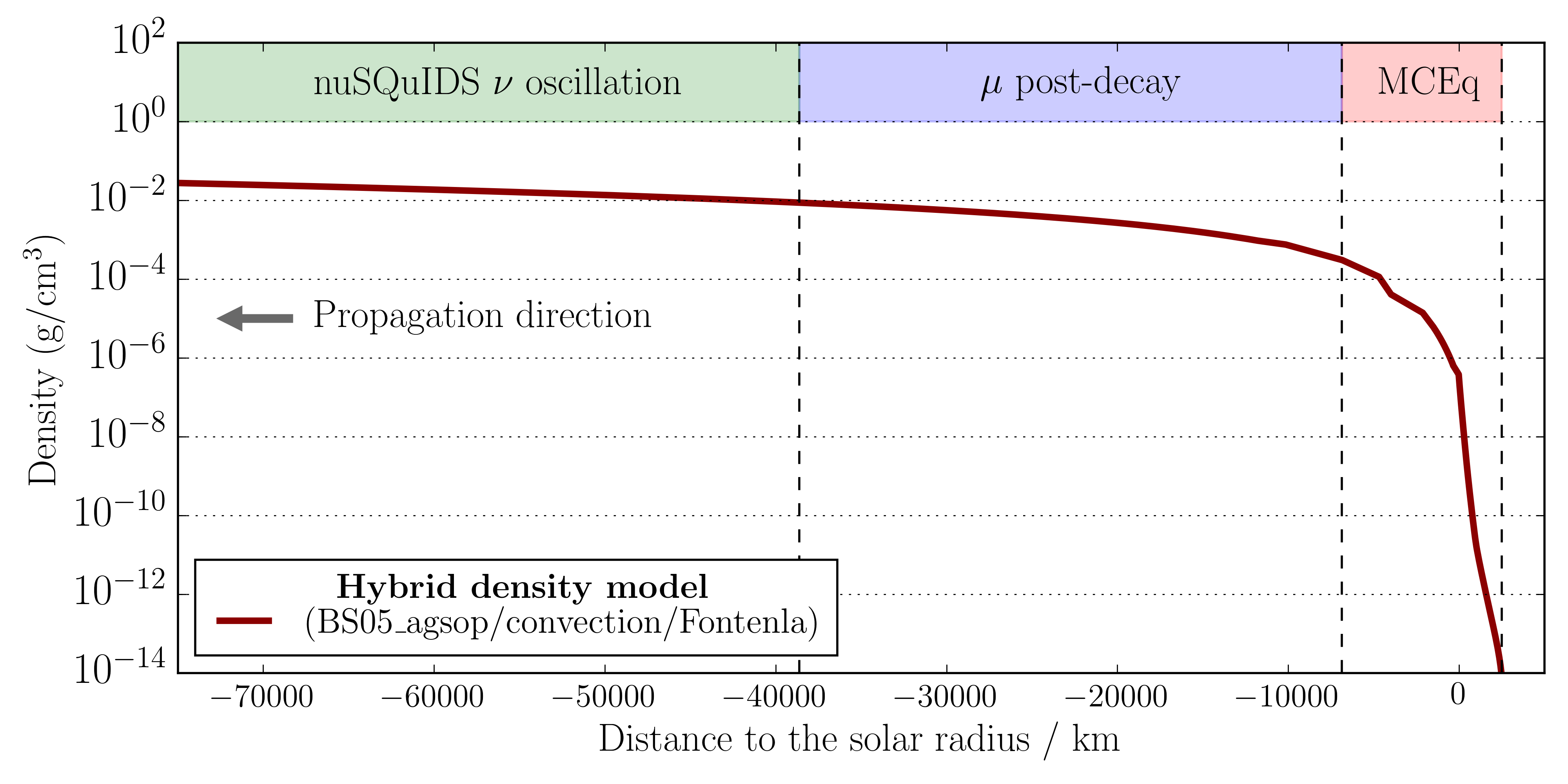}
\includegraphics[width=0.9\textwidth]{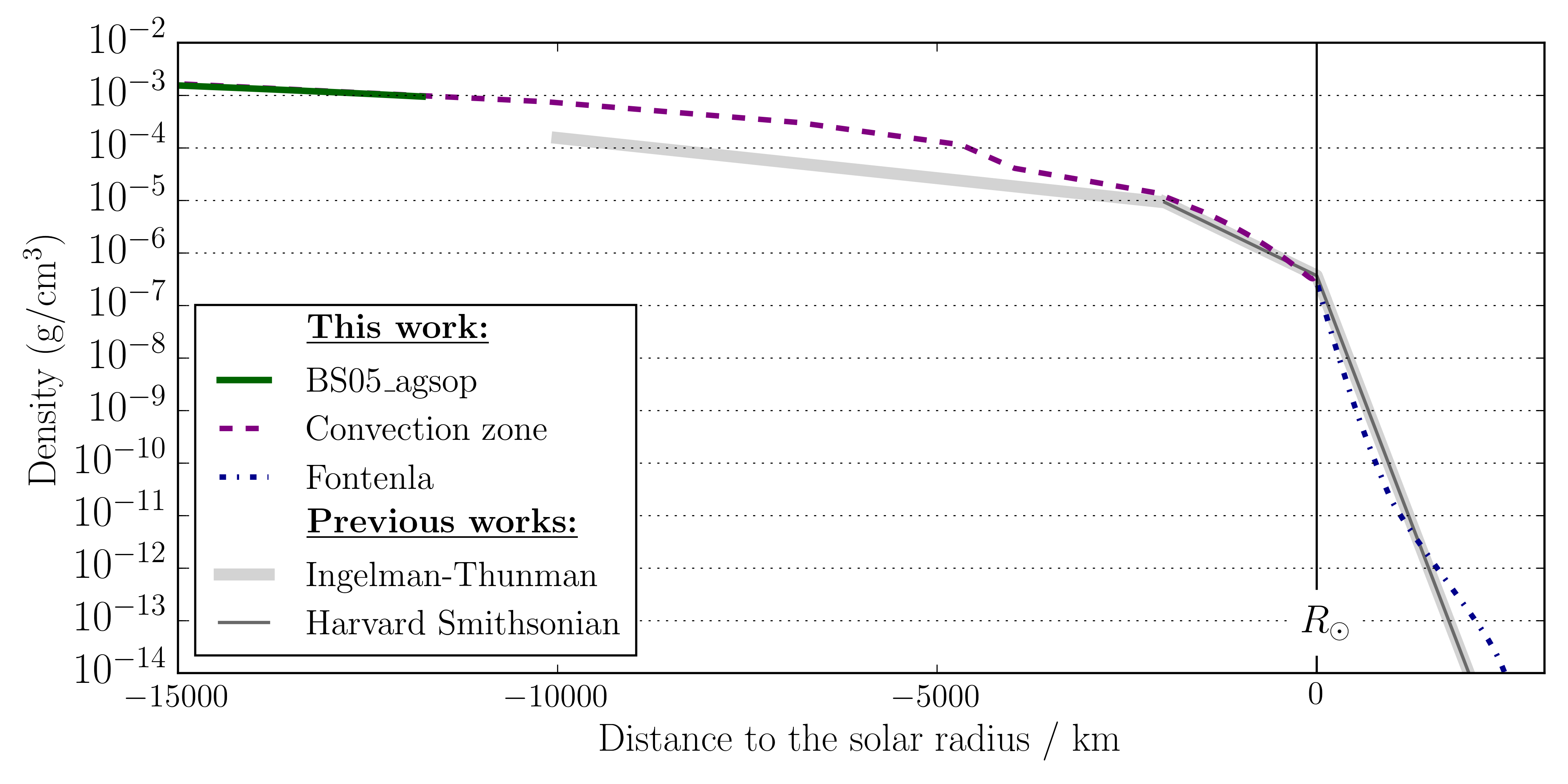}
\par\end{centering}

\caption{Top: Density profile in our hybrid solar density model, showing which radial regions the various stages of our calculation are performed. Bottom: deconstruction of our hybrid model in terms of its constituent parts, from (~\cite{vernazza,fontenla,bs05}). Shown for comparison is the model used in earlier flux calculations by Ingelman and Thunman~\cite{Ingelman:1996mj}. Note that although we briefly use a part of the Harvard-Smithsonian model~\cite{harvard} to patch a gap between the atmosphere and convection zone, this region is too short to highlight effectively on the above plot.\label{fig:SolarDensityProfile}}
\end{figure}

One of our motivations to improve on the solar neutrino flux calculations presented in past works (e.g. Ingelman and Thunman~\cite{Ingelman:1996mj}), was the significant improvements made on solar atmosphere modelling in the past years. Ingelman and Thunman used an exponential density profile fit on  the data of~\cite{vernazza} for the atmosphere and of~\cite{christensen} for the deeper layers of the Sun.
An update of~\cite{vernazza} has been produced by Fontenla et al.~\cite{fontenla}. As its predecessor, this model is a 1D semi-empirical model of the quiet-Sun. A 3D model (e.g.~\cite{asplund}) including the movements of the convection cells is not needed in the work presented here as the speed of these cells ($\sim$ 1km/s) can be neglected compared to the speed of the particles we consider.

An additional requirement for our solar model is a continuous density profile from the top of the atmosphere to the interior of the Sun. While the interaction region will be located in the low chromosphere and the photosphere, the density and composition of the core will be of importance for the neutrino propagation across the Sun.

A hybrid model based on the following combination allows to fulfill the two requirements described above:
\begin{itemize}
\item An up-to-date 1D atmosphere model~\cite{fontenla},
\item A brief use of the Harvard-Smithsonian atmosphere model~\cite{harvard}, extensively used in high-energy physics, for the deeper layers of the atmosphere,
\item A convection zone modeled by~\cite{spruit}. This maintains continuity between the deep layers of the atmosphere and the core. 
\item A core based on the BS05\_agsop model~\cite{bs05}.
\end{itemize}

The complete hybrid model density profile is shown in Figure~\ref{fig:SolarDensityProfile}, along with the delineation of the zones of our calculation: first {\tt MCEq} air-shower cascade evolution in high solar atmosphere, followed by numerical muon decay simulation in the convection zone, and finally {\tt nuSQuIDS} neutrino density matrix evolution across the core to the opposite side of the Sun.  Figure~\ref{fig:SolarDensityProfile}, bottom shows how our hybrid model is assembled in terms of the parts described above, and also how it compares to the previous approximation for the solar density profile used in~\cite{Ingelman:1996mj}.  
 
In this calculation we follow previous works such as~\cite{Ingelman:1996mj} by considering neutrino production in the atmosphere of a non-magnetic Sun.  It has been suggested~\cite{Seckel:1991ffa} that at the lowest cosmic ray energies, corrections to the neutrino flux due to various atmospheric and heliospheric effects may be non-negligible. Among these effects, the most relevant in the current framework are the propagation of low energy cosmic-ray through large-scale, such as the interplanetary magnetic field or the coronal field, and small-scale magnetic fields taking place in the low atmosphere of the Sun~\cite{Seckel:1991ffa}. The large scale fields will reduce the fraction of the cosmic ray flux that are absorbed by the Sun, leading to a smaller neutrino yield from the low energy cosmic rays. The latter changes the cascade evolution through charged particle reflection from solar magnetic flux tubes anchored in the bottom of the photosphere. The cosmic ray particle, or one of its subsequent charged products, can be mirrored and trapped by the magnetic structure, modifying the density profile seen by the cascade.
Because of the complexity of accurately modelling the effects of solar magnetic fields, such corrections are not included in the present work. This may introduce some additional uncertainty at the lowest energies (below 100~GeV) though over most of the energy range of our calculation the effects are expected to be sub-dominant to other sources of uncertainty.

\subsection{Numerical cascade equation solution in the solar atmosphere \label{sec:MCEq}}

The software package {\tt MCEq}~\cite{Fedynitch:2015zbe,Fedynitch:2015zma} is a linear, one-dimensional cascade equation solver. We use it to calculate the production neutrino fluxes from hadrons and muons produced in cosmic ray induced showers in the solar atmosphere.

\begin{figure}[t]
\begin{centering}
\includegraphics[width=0.99\textwidth]{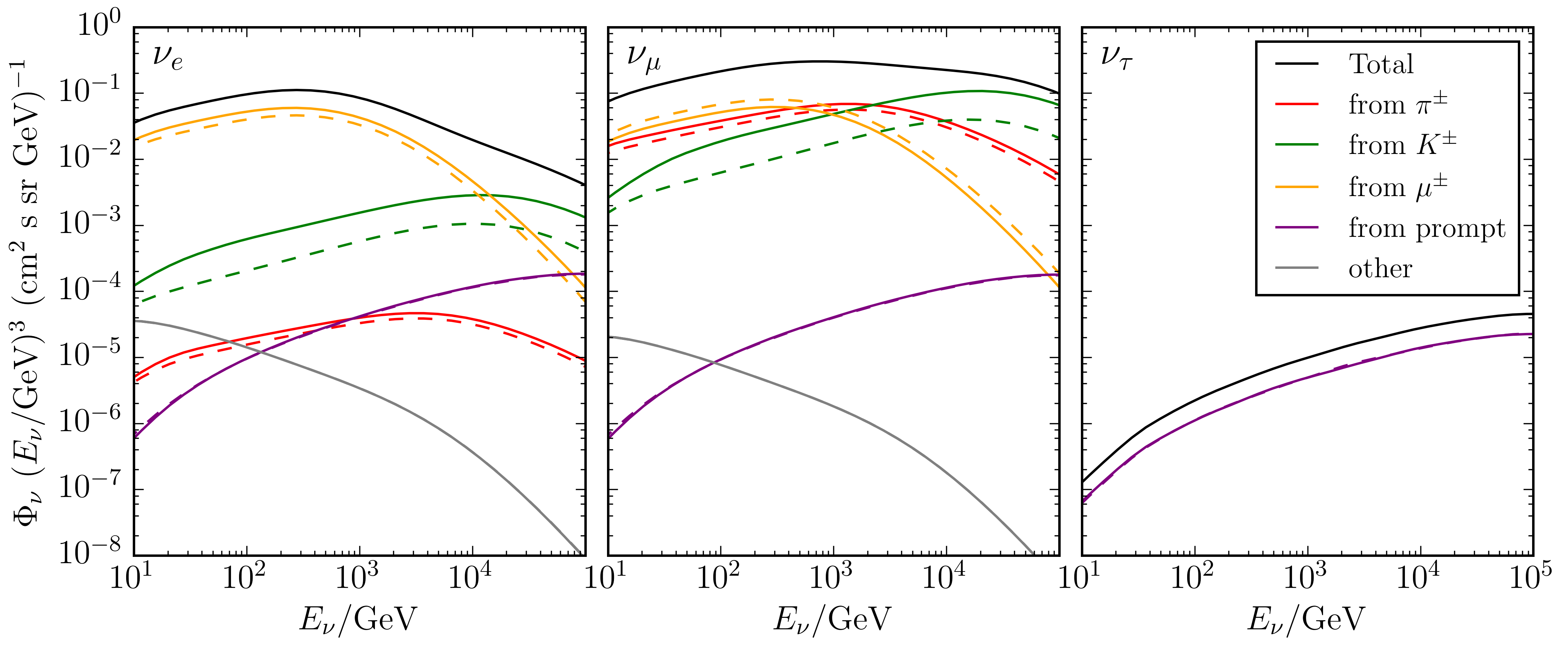}
\par\end{centering}

\caption{Contributions to the solar atmospheric neutrino flux at production for each neutrino flavor for impact parameter b=0. Solid lines show the contribution from neutrinos, dashed lines from antineutrinos. The black line shows the total the sum of neutrinos and antineutrinos of each flavor. \label{fig:FluxCompositionCentralModel}}
\end{figure}

We consider profiles at different cosmic ray impact parameters across the solar sphere, and these trajectories define chords through the solar atmosphere. Density profiles along the chords are calculated at discrete impact parameter values of $0<b<1$, with $b=0$ corresponding to a directly core-crossing trajectory and $b=1$ a glancing trajectory at the solar radius. Our calculation is divided into distinct regions as a function of radial depth.  The cascade equation for neutrino production is solved for each chord to predict the total flux crossing a surface approximately 6500 km inside the conventionally defined solar radius (see Figure~\ref{fig:SolarDensityProfile}).  The section of solar density profile used for the cascade calculation thus starts at a solar density of $10^{-14}\,\mathrm{g\,cm}^{-2}$ and ends at $3\times 10^{-3}\,\mathrm{g\,cm}^{-2}$.  Shorter geometries were also tested with negligible difference observed in the final $\pi$- and $K$-induced neutrino flux, validating that neutrino production by hadrons is complete by this radius and in our energy range. The most important region for neutrino production in the solar atmosphere lies at densities of $1\times 10^{-4} \mathrm{g\,cm^{-3}}$. This reflects that distance scales of the solar atmosphere are sufficiently long that by this point there has been ample column depth for the initial flux of primaries to have fully interacted, and for the resultant boosted hadrons to decay. 

\begin{figure}[t]
\begin{centering}
\includegraphics[width=0.99\textwidth]{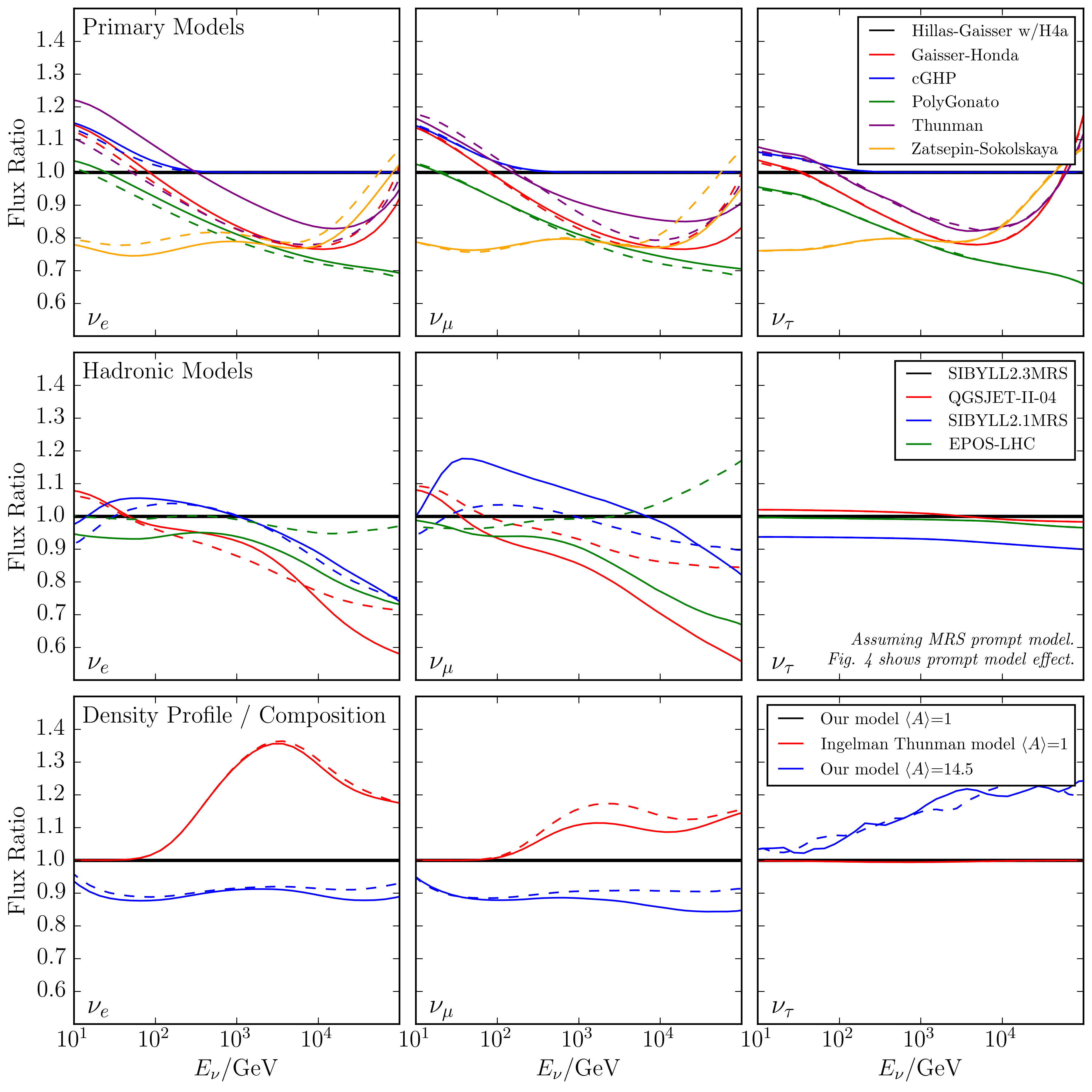}
\par\end{centering}

\caption{Effects of different models on our flux prediction, for impact parameter b=0. The top row shows various primary models; the second row, hadronic and composition models; the third row, extremal solar density and composition models. See text for more information and references.\label{fig:SystematicsPanel}}
\end{figure}

An exception to the above principle is the large flux of highly boosted muons, many of which remain at the end of the {\tt MCEq} simulation region, and have to be decayed subsequently in the next region.  After the {\tt MCEq} calculation is complete, the flux of muons penetrating out of the cascade equation geometry is extracted and fed to a second post-decay simulation step, which propagates them through a further radial 28000~km of solar atmosphere, to a region with density around $10^{-2}\,\mathrm{g\,cm^{-3}}$.  This region has sufficient integrated density that the total energy loss of the most energetic muons generated in the MCEq stage moves them below the lowest energy bin.  In each bin of the calculated muon flux, energy losses are applied according to tables from~\cite{Groom:2001kq} in steps, and at each step a fraction of the muon flux is converted into a neutrino population via three-body decay according to the polynomial parametrization from ~\cite{Lipari:1993hd}.  The propagation is evaluated in distance steps corresponding to whichever is the smallest between 1\% of the decay length of the boosted muon and 1\% of the energy loss length, and repeated until the muon energy falls below the threshold of 10~GeV relevant to this study. The effect of the additional post-decayed neutrino flux is most relevant for the electron flavor neutrinos, which are dominated over most of the energy range of our study by production from muon decay.  At energies below 100~GeV, this post-decayed flux makes a small ($<3$\%) contribution, since at low energies most muons have decayed in the {\tt MCEq} volume.  As the energy increases and the muons become more boosted, the additional flux becomes a more substantial component, contributing an additional 40\% to the electron neutrino flux and 9\% to the muon neutrino flux at 1~TeV. At still higher energies, the relative flux of the re-decayed muons falls again as muons become subdominant to hadrons like $\pi$, $K$, and eventually charmed hadrons.

The predicted flux of $\nu_e$, $\nu_\mu$, and $\nu_\tau$ in the solar atmosphere after both {\tt MCEq} and post-decay steps is shown in Figure~\ref{fig:FluxCompositionCentralModel}.  For illustration,  the results at impact parameter $b=0$ are shown, although the impact parameter dependence of the flux is weak at this stage. To obtain the total flux from the whole Sun, the flux at each $b$ must be propagated across the solar body, which introduces much larger b dependencies that must be accounted for before integrating. This step will be discussed in Section~\ref{sec:nuSQUIDS}.

In Figure~\ref{fig:FluxCompositionCentralModel}, the total flux, shown as a black line, is broken into neutrino and antineutrino components, and separated by parent particle. The electron neutrino / antineutrino component is dominated by muon decay until the highest energies where this becomes subleading to kaon decays. The muon neutrinos / antineutrinos are dominated by muon decays at the lowest energies, then briefly by pions at around 1~TeV, and finally by kaon decay at the highest energies.  Neutrinos from charmed hadrons, often called ``prompt'' neutrinos, are sub-dominant everywhere for the muon and electron flavors. However, since neither pions, kaons or muons have sufficient mass to produce a $\tau$ lepton, this charmed contribution is the only source of tau neutrinos, as can be seen in Figure~\ref{fig:FluxCompositionCentralModel}, right. 

The inputs to the calculation described above are:

\begin{enumerate}[I]%for capital roman numbers.
\item The solar density profile, as discussed in Section~\ref{sec:SolarAtmopshere}.
\item The composition / hadronic interaction model, which are supplied together as {\tt MCEq} yield and decay tables, used in solving the cascade equation. For the calculation shown in Figure \ref{fig:FluxCompositionCentralModel} we used the {\sc SIBYLL}-2.3~\cite{Engel:2015dxa,Riehn:2015oba} model assuming a purely protonic atmosphere and {\sc MRS} prompt model~\cite{Martin:2003us}.
\item  The primary cosmic ray spectrum, which is injected as the initial condition.  For the calculation shown in Figure~\ref{fig:FluxCompositionCentralModel} we used the cosmic ray flux model H4a from \cite{Gaisser:2011cc}, also called the Hillas-Gaisser H4a model.
\end{enumerate}

To assess the scale of systematic uncertainty on our prediction we explore a variety of primary models and composition / hadronic models. In addition to the H4a parametrization of~\cite{Gaisser:2011cc}, the primary models studied include the ``Gaisser-Honda'' model, a widely used five-mass-group model tuned to balloon data~\cite{Gaisser:2002jj}; the ``combined Gaisser-Honda with H4a'' model ({\sc cHGP}) \cite{Fedynitch:2012fs};  the ``Thunman model'', a broken power law parametrization constructed in \cite{Gondolo:1995fq}; the ``Polygonato model''~\cite{Hoerandel:2002yg}, constructed to explain the knee of the cosmic ray spectrum in terms of sequential cut-offs of different mass components; and the ``Zatsepin-Sokolskaya'' model~\cite{Zatsepin:2006ci}, constructed in terms of populations of cosmic rays accelerated by discrete types of nova and supernova events.   The effects of varying the primary model on each of the flavor fluxes in the lower solar atmosphere, relative to the prediction in Figure \ref{fig:FluxCompositionCentralModel} is shown in the top row of Figure~\ref{fig:SystematicsPanel}.

\begin{figure}[t]
\begin{centering}
\includegraphics[width=0.99\textwidth]{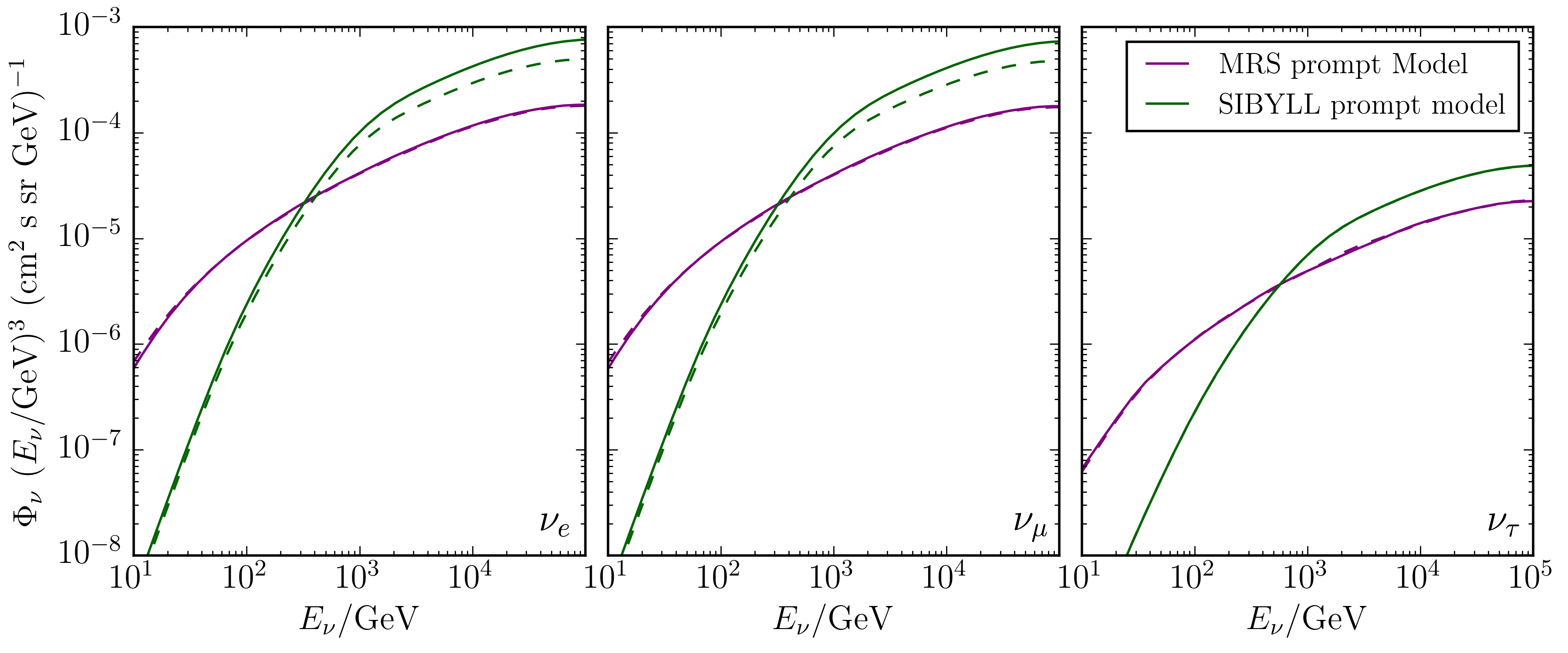}
\par\end{centering}
\caption{The prompt neutrino flux from two different charm models ({\sc MRS}~\cite{Martin:2003us} and {\sc SIBYLL}-2.3~\cite{Engel:2015dxa,Riehn:2015oba}),for impact parameter b=0.\label{fig:PromptPlot}}
\end{figure}

The set of hadronic models we consider are
{\sc SIBYLL}-2.3~\cite{Engel:2015dxa,Riehn:2015oba}; the previous {\sc SIBYLL} version, {\sc SIBYLL}-2.1 \cite{Ahn:2009wx};  the {\sc EPOS-LHC}~\cite{Pierog:2013ria} model; and {\sc QGSJET-II}-04 \cite{Ostapchenko:2013pia,PhysRevD.83.014018}.  In each case we couple the model for conventional hadron production to the {\sc MRS} prompt model~\cite{Martin:2003us}. With the exception of {\sc SIBYLL}-2.3, these models do not have {\sc MCEq} tables prepared for purely protonic environments. Hence to estimate the effect of changing the hadronic model we take the ratio of the flux predictions calculated using an air atmosphere of the same nucleon density profile as the Sun, $\phi_{air}^{model}(E)$.  We then correct this flux using the {\sc SIBYLL}-2.3 flux predicted in a protonic atmosphere according to the approximation $\phi_{p}^{model}(E)\approx\phi_{air}^{model}(E)/\phi_{air}^{SIBYLL2.3}(E)\times\phi_{p}^{SIBYLL2.3}(E)$. The effects of varying the hadronic model is shown in the central panel of Figure~\ref{fig:SystematicsPanel}.

Although we use the {\sc MRS} model in all cases above, the {\sc SIBYLL} models also have a built-in charm production parametrization available. We compare the effects of switching to this charm model in Figure~\ref{fig:PromptPlot}. This comparison is made using {\sc SIBYLL}-2.3 coupled to either its intrinsic charm model or to the MRS model, in both cases using a protonic atmosphere.  The charm model uncertainties have a drastic effect on the prompt flux prediction, though this is a small effect for the $\nu_\mu$ and $\nu_e$ fluxes where the prompt contribution is strongly sub-dominant.  On the other hand, $\nu_\tau$ are produced exclusively through decays of charmed hadrons in this energy range, so the charm model effects are the leading source of uncertainty at production for $\nu_\tau$ in the solar atmosphere. The scale of the variation induced by changing the charm model on the prompt fluxes is shown in Figure~\ref{fig:PromptPlot}.

Finally we investigate the effects of changing the atmospheric density and composition. We find that extreme outliers are required to produce flux distortions comparable with the hadronic and primary model uncertainties. Figure~\ref{fig:SystematicsPanel}, bottom shows a comparison between an assumed protonic ($
\langle A \rangle = 1$) or air ($
\langle A \rangle = 14.5$) composition for the solar atmosphere.  With this highly unrealistic composition model, an order 10\% distortion is observed.  Similarly, switching to the much less accurate Ingelman Thunman atmospheric density model allows us to test the effect of a gross distortion in the solar density profile~\cite{Ingelman:1996mj}. Again, with such a highly exaggerated distortion, the flux variations are of comparable scale to those obtained from switching between primary and hadronic models. Because the true uncertainties on the composition and density profile are much smaller than those investigated here, which generate effects of comparable scale to the primary and hadronic model variations, we conclude that these uncertainties can be neglected in our final uncertainty budget.

\begin{figure}[t]
\begin{centering}
\includegraphics[width=0.8\textwidth]{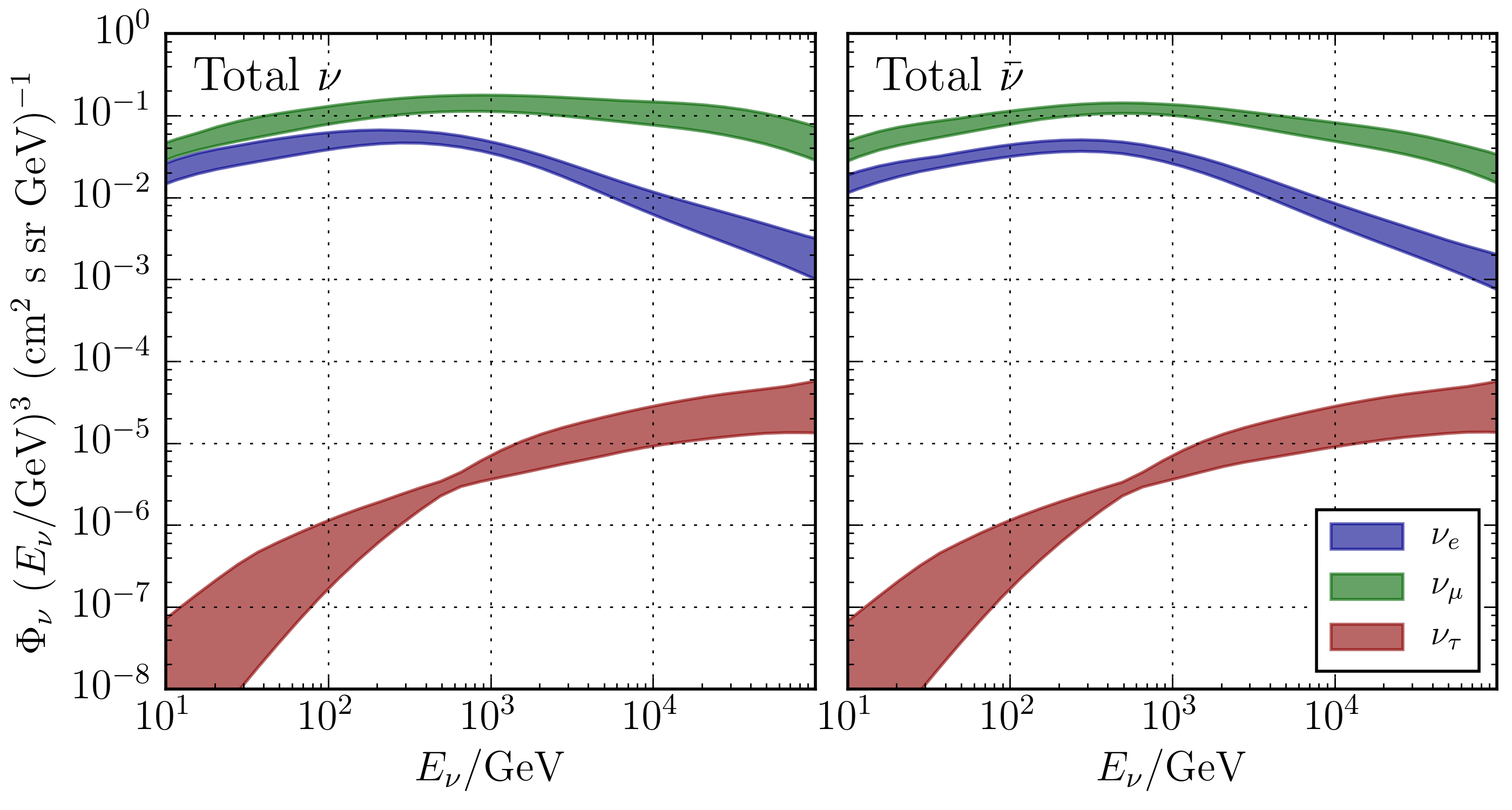}
\par\end{centering}
\caption{Total flux of $\nu_e$,$\nu_\mu$,$\nu_\tau$ produced in solar showers before propagation effects for impact parameter b=0, with uncertainty bands from hadronic model, prompt model and primary model.\label{fig:FluxBands}}
\end{figure}

We compile the effects of all variations described above into a total uncertainty band for each neutrino and antineutrino flavor, shown in Figure~\ref{fig:FluxBands}.  The band is drawn between a flux evaluated at the extremal interaction model, charm model, and primary model combination at each energy.  We conclude that the $\nu_\mu$ and $\nu_e$ fluxes in both neutrinos and antineutrinos are predictable within a factor of approximately 2 below 1~TeV, with slightly larger uncertainties at higher energies.  The $\nu_\tau$ flux is dominated by charm production uncertainties and is much less predicable than $\nu_\mu$ or $\nu_e$.  While the $\nu_\tau$ flux at production is difficult to estimate, as will be shown in the next section, oscillations during solar propagation strongly mix the $\nu_\mu$ and $\nu_\tau$ flavors, resulting in the charm model uncertainty not being significant in our final results.

\subsection{Oscillation and absorption effects \label{sec:nuSQUIDS}}

\begin{figure}[t]
\begin{centering}
\includegraphics[width=\textwidth]{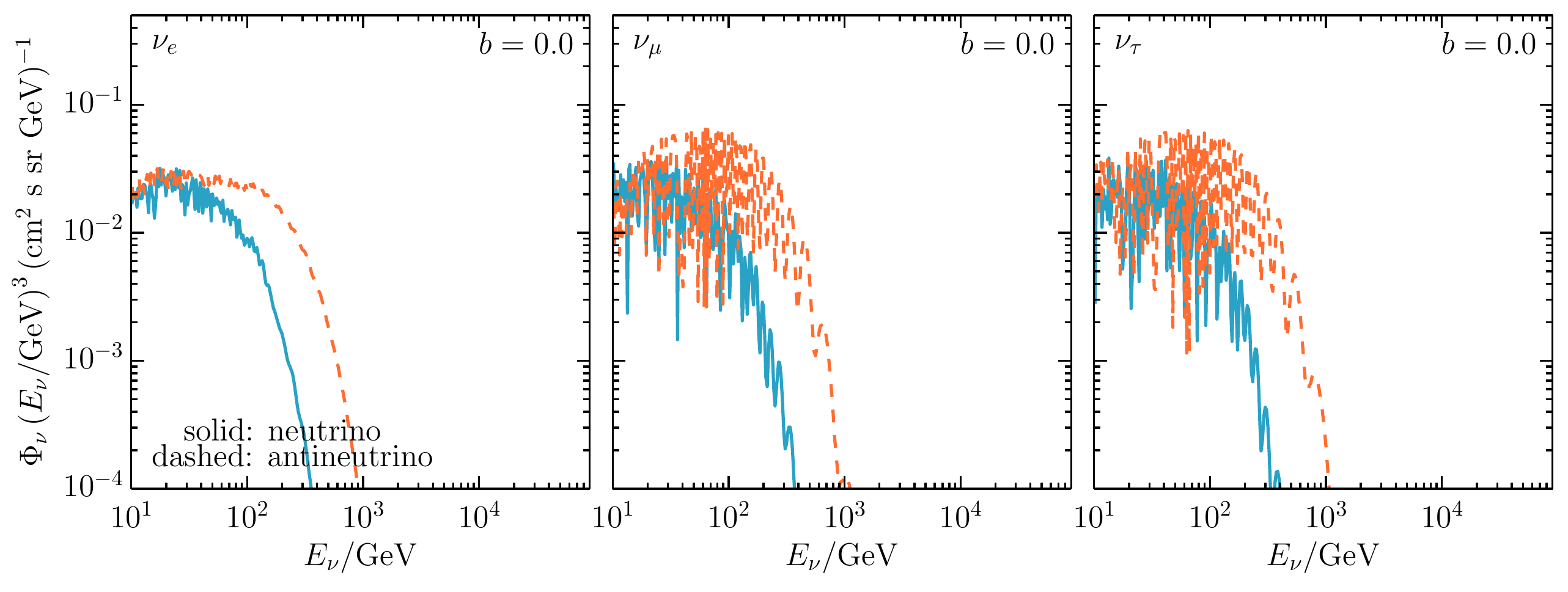}\\
\includegraphics[width=\textwidth]{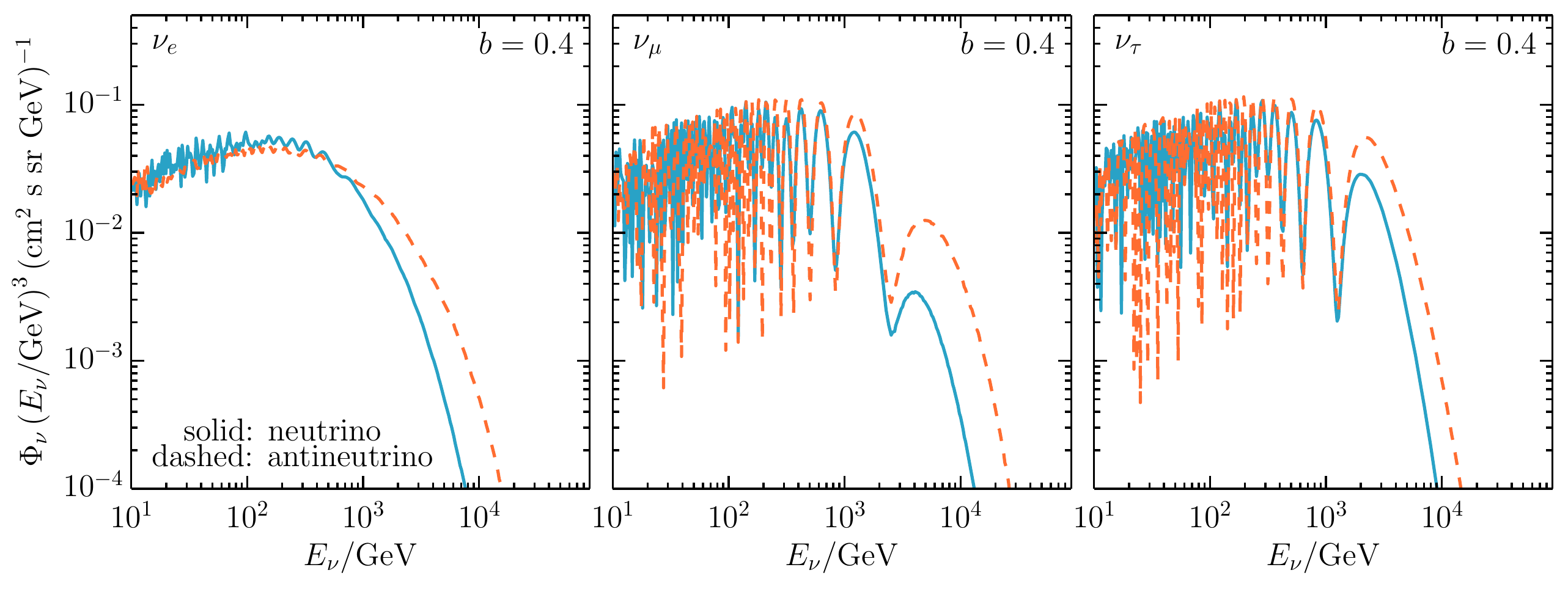}\\
\includegraphics[width=\textwidth]{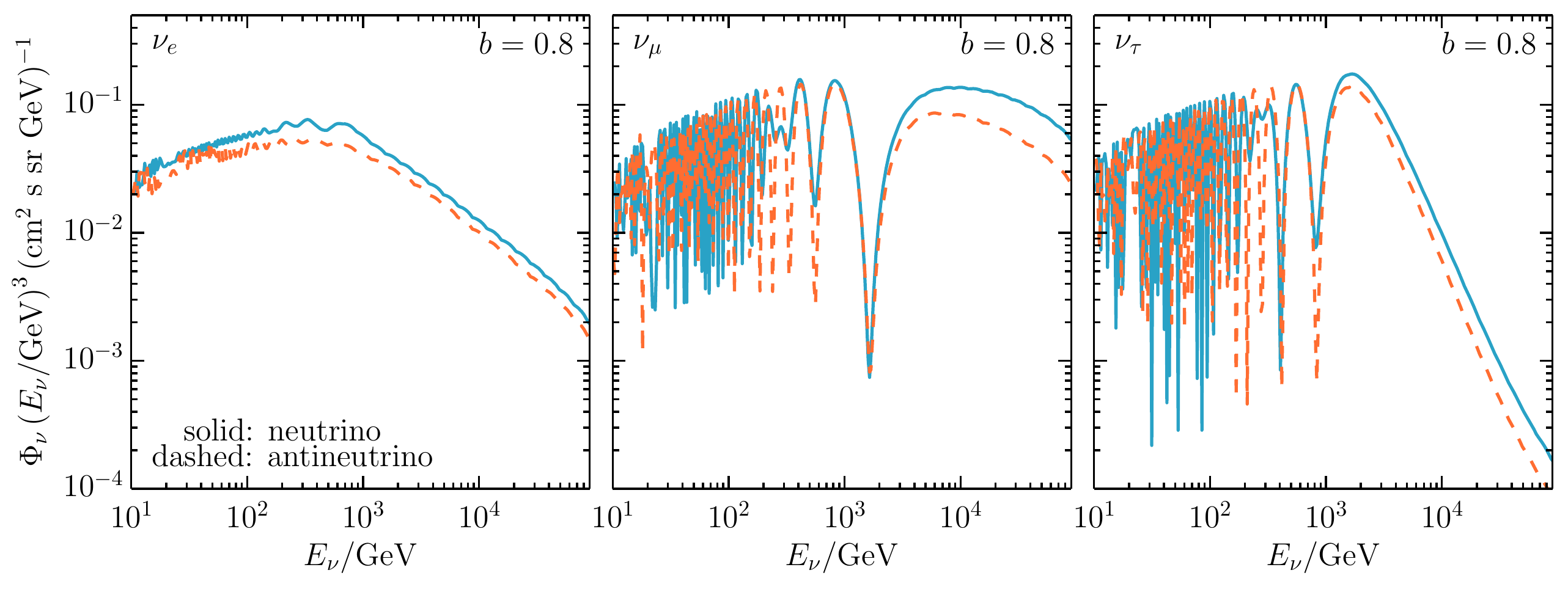}
\par\end{centering}
\caption{Each panel shows the neutrino flux per flavor as solid light blue lines and the corresponding antineutrino flux as bright orange dashed lines. Propagated fluxes are shown for three different impact parameters: 0.0 (upper panel), 0.4 (middle panel), and 0.8 (lower panel).}
\label{fig:propagated_fluxes}
\end{figure}

The neutrino flux calculated in Section~\ref{sec:MCEq} at the outer most layers of the Sun needs to be propagated through the dense solar medium. For each impact parameter, $b$, we assume a one dimensional propagation through the Sun. As neutrinos travel through it, two important phenomena will modify the flavor composition and flux strength: neutrino oscillations and solar absorption effects, respectively. To first order, the opacity induced by traversing the Sun is given by 
\begin{equation}
\tau(E_\nu;b) = {\rm exp}\left[-\int_{l(b)} n_N(x;b) \sigma_{\nu N}(E_\nu)  ~ dx\right],
\label{eq:absorption}
\end{equation}
where $l(b)$ is the chord traversed, $n_N(x;b)$ is the nucleon number density along the trajectory, and $\sigma_{\nu N}(E)$ is the total neutrino nucleon cross section. In the absorption the neutrino electron induced opacity is ignored since, for $10~ {\rm GeV} < E_\nu< 1~ {\rm PeV}$, $\sigma_{\nu e}(E) \ll \sigma_{\nu N}(E)$ \cite{Gandhi:1998ri,Formaggio:2013kya}. The nucleon number density is obtained from~\cite{bs05}. We use the neutrino nucleon cross section as calculated in~\cite{Arguelles:2015wba}. Equation \eqref{eq:absorption} has some deficiencies that prevent us from using it directly.  First, it neglects neutrino neutral current re-population and also tau regeneration, which make the Sun more transparent to high-energy neutrinos~\cite{Halzen:1998be,Dutta:2005yt}. Furthermore, it omits neutrino oscillations. A consistent formalism to treat both neutrino interactions together with oscillation, and including tau regeneration, is given in \cite{Sigl:1992fn,GonzalezGarcia:2005xw,Salvado:2016dsf}; which relies on the density matrix formalism to represent the neutrino flux and describe the neutrino evolution by a quantum Boltzmann equation. This formalism is implemented in the publicly available {\tt nuSQuIDS} package~\cite{Delgado:2014kpa,nusquids}, which is used in this work. We set the neutrino propagation medium according to the solar model given in~\cite{bs05} and set the neutrino oscillation parameters to the best fit values of \cite{Esteban:2016qun}. This implementation has been tested by reproducing similar propagation calculations shown in~\cite{Cirelli:2005gh,Blennow:2007tw,Arguelles:2012cf}.

The result of the neutrino propagation through the Sun is shown in Figure~\ref{fig:propagated_fluxes} for three different impact parameters. As expected due to oscillations the $\nu_\mu$ and $\nu_\tau$ fluxes are at similar intensity, where as oscillations with the $\nu_e$ flavor are suppressed by matter effects in the inner part of the Sun~\cite{Wolfenstein:1977ue,Mikheev:1986wj,Giunti:2009xz}. At 1 TeV the matter suppression is significant until approximately $0.9 R_\odot$, whereas at a 100 GeV its only dominant until $\sim 0.5 R_\odot$. When matter effects are no longer dominant the oscillation scale is comparable to the solar radius at 100 {\rm GeV} and $\sim$10\% at 1 TeV for $\nu_e\to\nu_x$ transitions.  Furthermore, the solar opacity increases with energy and is larger for neutrinos than antineutrinos.

\subsection{Predicted solar atmospheric neutrino flux at Earth \label{sec:AtEarth}}

\begin{figure}[t]
\begin{centering}
\includegraphics[width=\textwidth]{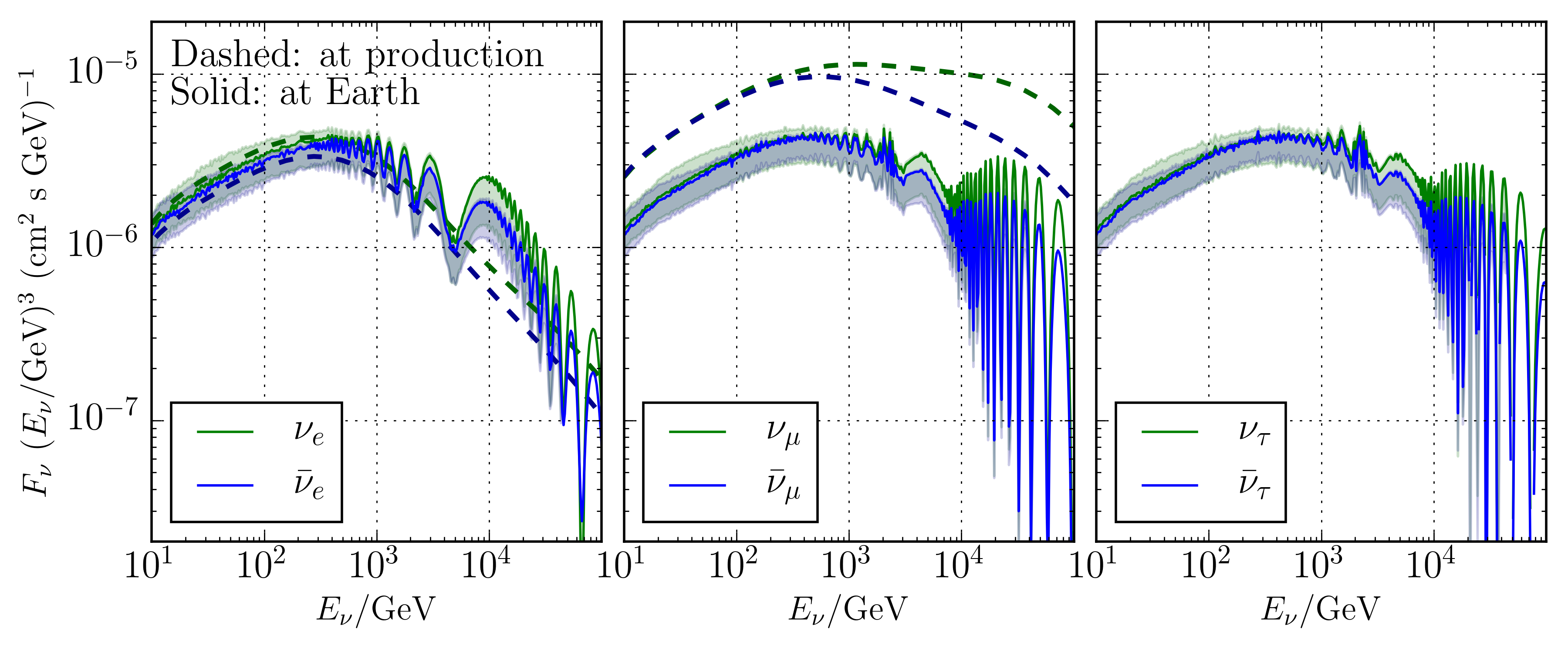}
\par\end{centering}
\caption{The fluxes arriving at Earth (solid) compared with those produced in the solar atmosphere (dashed).  The lines show the result for a specific shower model (H4a, {\sc SIBYLL}-2.3 interaction model and MRS prompt).  The bands show the uncertainty region across all models.  The three panels show the flux for each neutrino flavor, in both neutrinos and antineutrinos.}
\label{fig:aggregated_fluxes}
\end{figure}

In order to predict the flux at the Earth from the results of Section \ref{sec:nuSQUIDS} we must aggregate the neutrino fluxes calculated at several impact parameters, and also account for oscillation effects between Sun and Earth. The aggregated atmospheric solar neutrino flux is given by 
\begin{equation}
\Phi(E_\nu)_\alpha =  2 \int db~ b \Phi_{b,\alpha}(E_\nu) \Omega_\odot,
\label{eq:flux_combination}
\end{equation}
where $\phi_{b,\alpha}(E_\nu)$ is the flux of flavor $\alpha$ at impact parameter $b$ after solar propagation and $\Omega_\odot$ is the total solid angle of the Sun. The aggregated flux is dominated by the larger impact parameters due to a geometrical enhancement and because the fluxes are less suppressed as $b$ increases (see Fig. \ref{fig:propagated_fluxes}).  We evaluate the above integral over $b$ using a trapezium rule implementation with support points at intervals of $\delta b$=0.1.  Some uncertainty is introduced by this discretization, particularly in the outer layers of the Sun where the $b$ dependence is largest and where most of the solid angle is contained. This was assessed by repeating the calculation with ten extra support points between $b=0.90$ and $b=0.99$ for one flux model.  The integration uncertainty was found to be $<5$\% below 10~TeV, growing to ~10\% at 100~TeV where the absorption effects are largest, and thus sub-dominant to other uncertainties treated in our calculation.  

To account for neutrino oscillations from the Sun to the Earth we average the flux over one orbit, namely,
\begin{equation}
\bar\Phi_\alpha(E_\nu) =  \frac{1}{T}\sum_\beta \int_0^{T} dt~ P^{\beta,\alpha}_{osc}(r(t),E_\nu)\Phi_{b,\beta}(E_\nu),
\end{equation}
where $T$ is one year, $t$ is the time through the year, $r(t)$ is the distance from the Sun to the Earth, and $P_{osc}$ is the vacuum oscillation probability. The distance difference between the aphelion and perihelion is $4 \times 10^6 \,{\rm  km}$. This distance is $\sim 100\times$ larger than the neutrino production region (see Fig.~\ref{fig:SolarDensityProfile}), which implies we can ignore effects of oscillation averaged due to the production region uncertainties. The neutrino oscillation length in vacuum at $E_\nu=100~ {\rm GeV}$ is approximately 1\% of the Earth-Sun distance yearly variation, while they are comparable at 10 TeV. Thus neutrinos with energies less than $\sim 1$ TeV will exhibit oscillations which are averaged by the changing Earth-Sun distance, while oscillations at the highest neutrino energies remain observable. The result of this process is shown in Fig.~\ref{fig:aggregated_fluxes}.

\section{Solar atmospheric neutrinos as a signal\label{sec:Signal}}

\begin{figure}[t]
\begin{centering}
\includegraphics[width=0.9\textwidth]{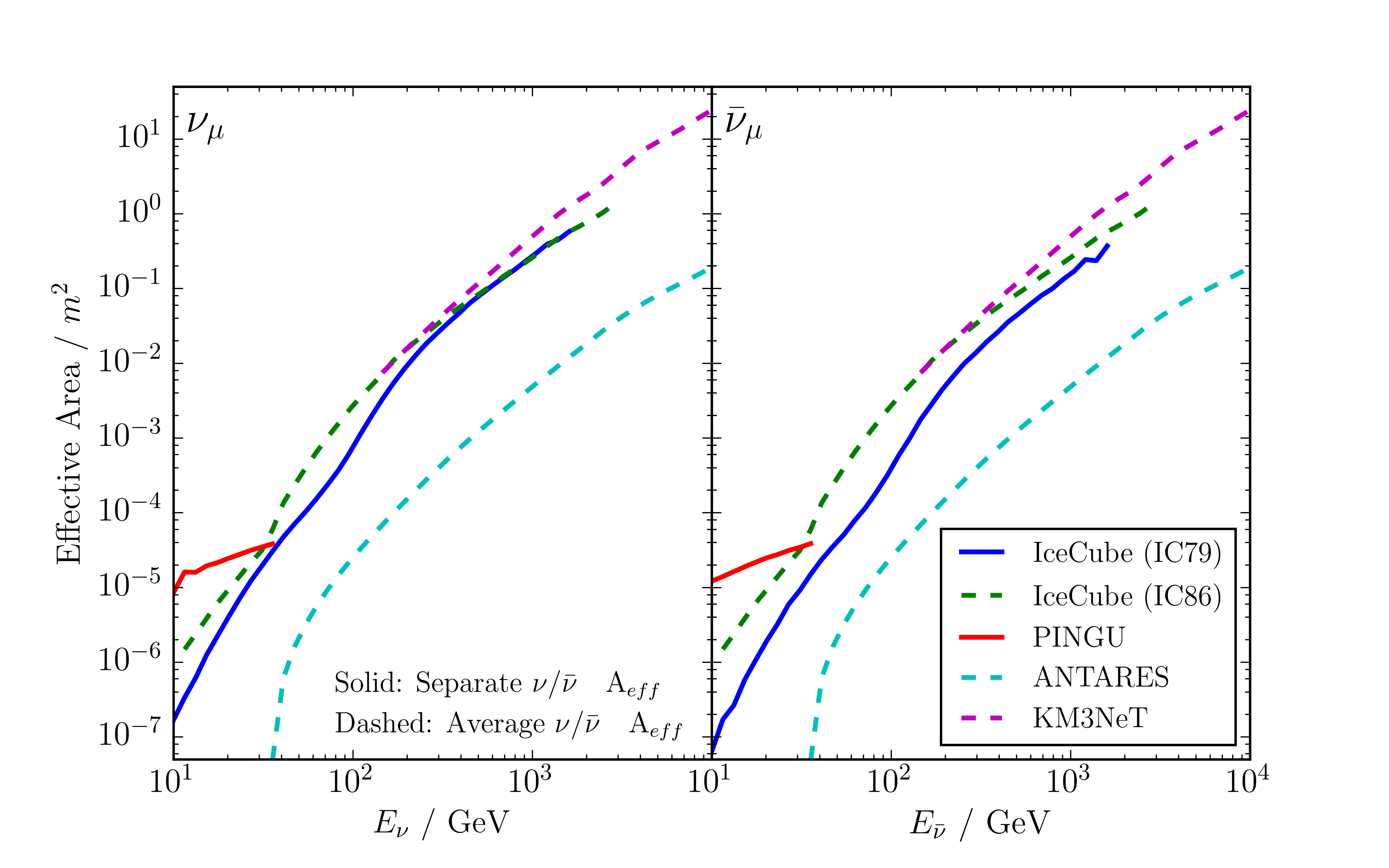}
\par\end{centering}

\caption{Effective areas of various neutrino detectors considered in this study for $\nu_\mu$ (left) and $\bar\nu_\mu$ (right). The solid lines represent effective areas that are provided separately for $\nu_\mu$ and $\bar\nu_\mu$.  The dashed lines are effective areas that are provided averages over $\nu_\mu$/$\bar\nu_\mu$, and so are the same on the left and right panels. \label{fig:EffectiveAreas}}
\end{figure}

Having predicted the solar atmospheric neutrino flux at Earth, it remains for us to consider its observability by realistic experiments.  To calculate the number of observable events and their spectrum, we require both a flux and a detector effective area. The effective areas we consider are from a range of existing and proposed neutrino telescopes, shown in Figure~\ref{fig:EffectiveAreas}.  In all cases, we consider only the $\nu_\mu$ / $\bar{\nu}_\mu$ channel, since only this sample is expected to have sufficient pointing resolution to perform a high quality point source search in the direction of the Sun. 

\begin{table}[b!]
\begin{centering}
\begin{tabular}{|c|c|c|}
\hline 
Experiment & Expected $\nu_{\mu}$ rate $R$ (evts / yr) & Expected $\bar{\nu}_{\mu}$ rate $R$ (evts / yr)\tabularnewline
\hline 
\hline 
IceCube (IC79) & $1.36 < R < 2.17$ & $0.73 < R < 1.17$\tabularnewline
\hline 
IceCube (IC86)* & $2.05 < R < 3.29$ & $1.97 < R < 3.16$\tabularnewline
\hline 
ANTARES* & $0.032 < R < 0.053$ & $0.030 < R < 0.049$\tabularnewline
\hline 
IceCube+PINGU & $1.42 < R < 2.26$ & $0.79 < R < 1.26$\tabularnewline
\hline 
KM3NeT* & $3.02 < R < 4.95$ & $2.78 < R < 4.53$\tabularnewline
\hline 
\end{tabular}
\par\end{centering}

\caption{The expected event rates in each of the detectors considered in this work. The rows marked with a * are for detector configurations where only the averaged effective area is provided, and so the neutrino and antineutrino effective areas set equal.  Because the better estimate is provided where the separate effective areas are available, the IceCube+PINGU rate is constructed using the IC79 and PINGU effective areas. \label{tab:ExpectedEventRates}}

\end{table}

The IceCube effective area is taken from a dedicated WIMP search analysis~\cite{Aartsen:2016exj} which has three discrete and independent sub-samples, and was provided in a data release available at~\cite{IC79DataRelease}.  We add the three effective areas to form a whole-experiment effective area for the purposes of this study. In the data release, effective areas are only provided up to an energy of 1775 GeV, and so we calculate the neutrino floor for IceCube-like effective areas up to this upper energy limit.  PINGU extends IceCube's reach to lower energies~\cite{Aartsen:2014oha}, and we assume that when PINGU runs, IceCube will also be running and maintaining its existing effective area at high-energy.  Thus we construct the combined effective area as the published PINGU effective area below 40 GeV joined onto the IceCube effective area at higher energies.  This is likely to be a slight underestimate of the PINGU+IceCube combined effective area since PINGU will contribute slightly to improving IceCube's sensitivity above 40 GeV. The more recent IceCube result~\cite{Aartsen:2016zhm} provides only effective areas averaged over neutrinos and antineutrinos, and so can only be included approximately in our calculations. Because the sensitivity floor is only weakly sensitive to the details of the effective area, we use the more complete effective area model of~\cite{IC79DataRelease} to calculate it, but will later estimate event rates for the new effective area from~\cite{Aartsen:2016zhm} by assuming equal effective areas for neutrinos and antineutrinos. Similarly, the ANTARES effective area reported in~\cite{ANTARESEffectiveArea} and the  KM3Net effective area from~\cite{Bagley:2009wwa} are both reported averaged over neutrinos and antineutrinos, and we make similar approximations in these cases.  The shape and location of the sensitivity floor, where solar atmospheric neutrinos overwhelm the expected dark matter signal, is independent of the normalization of the effective area and depends only on its shape.  We will later show that in all cases described, the location of the sensitivity floor is largely independent of the experiment under consideration.

The expected rate of detectable solar atmospheric neutrinos, on the other hand, does depend on the effective area magnitude as well as shape.  Considering the neutrino telescopes discussed above, we calculate the expected number of events per year detectable given each effective area. Our results are shown tabulated in Table~\ref{tab:ExpectedEventRates}.  If sufficient pointing resolution can be achieved, according to our predictions this flux should already be detectable above the terrestrial atmospheric neutrino background in IceCube's multi-year muon neutrino sample.

\section{Solar atmospheric neutrinos as a background\label{sec:Background}}

\begin{figure}[t]
\begin{centering}
\includegraphics[width=0.99\textwidth]{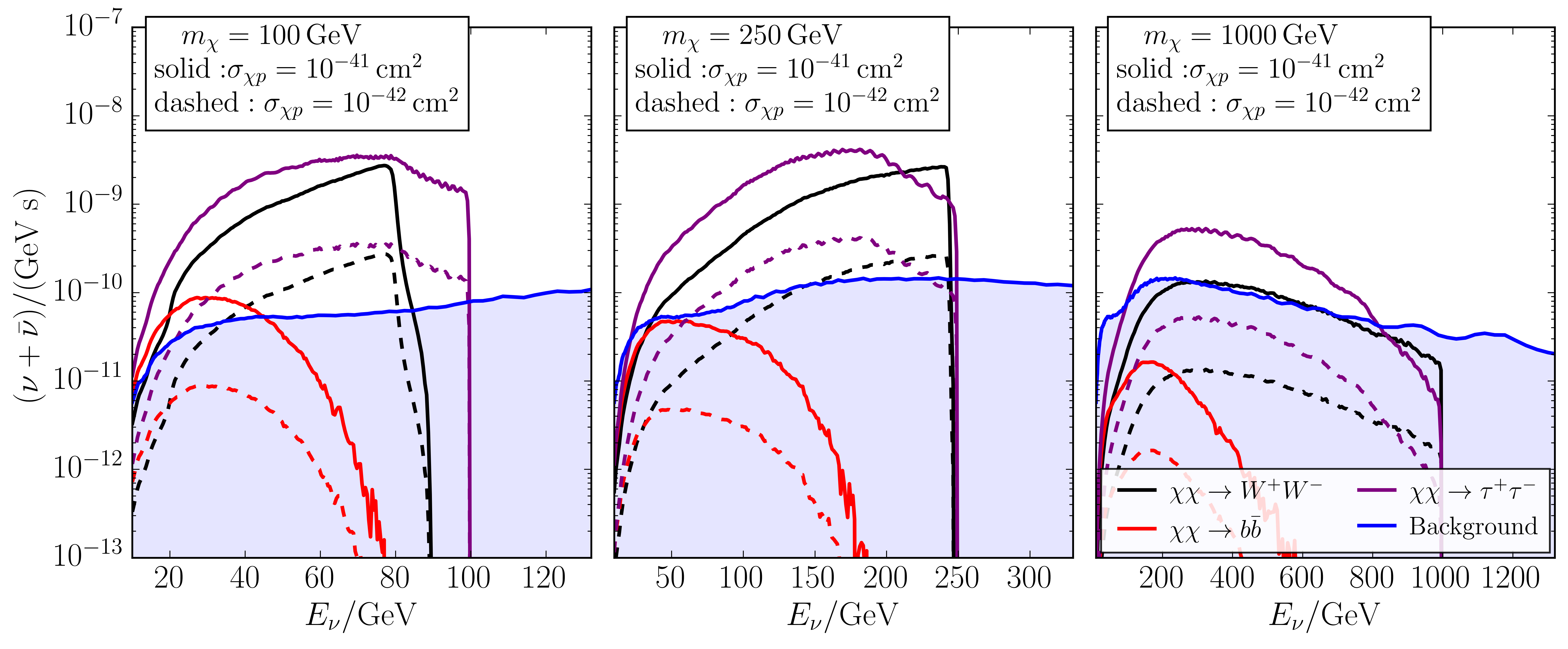}
\par\end{centering}

\caption{Comparison of the solar background to dark matter annihilation signals given three different masses and WIMP-proton spin-dependent cross sections.  For this comparison we use the IceCube effective area. \label{fig:IceCubeModels}}
\end{figure}

In this section we compare the detectable WIMP self-annihilation signals described in Section~\ref{sec:Signature}  with the background from the flux of high-energy, solar atmospheric neutrinos.  Figure~\ref{fig:IceCubeModels} shows the comparison between the detectable flux of dark matter from self-annihilation modes $\chi\chi\rightarrow b\bar{b}$, $\chi\chi\rightarrow W^{+}W^{-}$ and $\chi\chi\rightarrow \tau^{+}\tau^{-}$, for various values of the cross section near the sensitivity floor, and WIMP masses 100 GeV, 250 GeV and 1000 GeV. This plot uses the effective area of the IceCube detector from~\cite{IC79DataRelease}.  Our next goal is to establish quantitatively which regions of WIMP parameter space fall below the sensitivity floor imposed by the solar atmospheric neutrino background.

This task is non-trivial, given that there is no unique definition of the sensitivity floor.  We will define that the sensitivity floor has been reached when, in a representative toy analysis, the number of expected background events becomes equal in size to the signal being sought.  As in the case for the sensitivity floor in direct dark matter searches, the caveats that small signals can still be detected over large backgrounds with sufficiently large statistics, and that additional degrees of freedom can open up new parameter space, will also apply here.

We thus make the assumption that the detector is ideal in the sense that events can be identified as being of solar origin unambiguously through pointing information, and the only additional information is the event energy, which is known precisely. 
In reality, both direction and energy will suffer from imperfect reconstructions. The uncertainty on the neutrino energy, in particular, will depend on energy reconstruction of the detected muon as well as the distribution of the initial energy among interaction products. It will result in a smearing of the events around the true initial neutrino energy with a wider spread at low energies.
The neutrino floor derived in this work is therefore likely to be too deep, especially for low mass WIMPs.

\begin{figure}[t]
\begin{centering}
\includegraphics[width=0.35\textwidth]{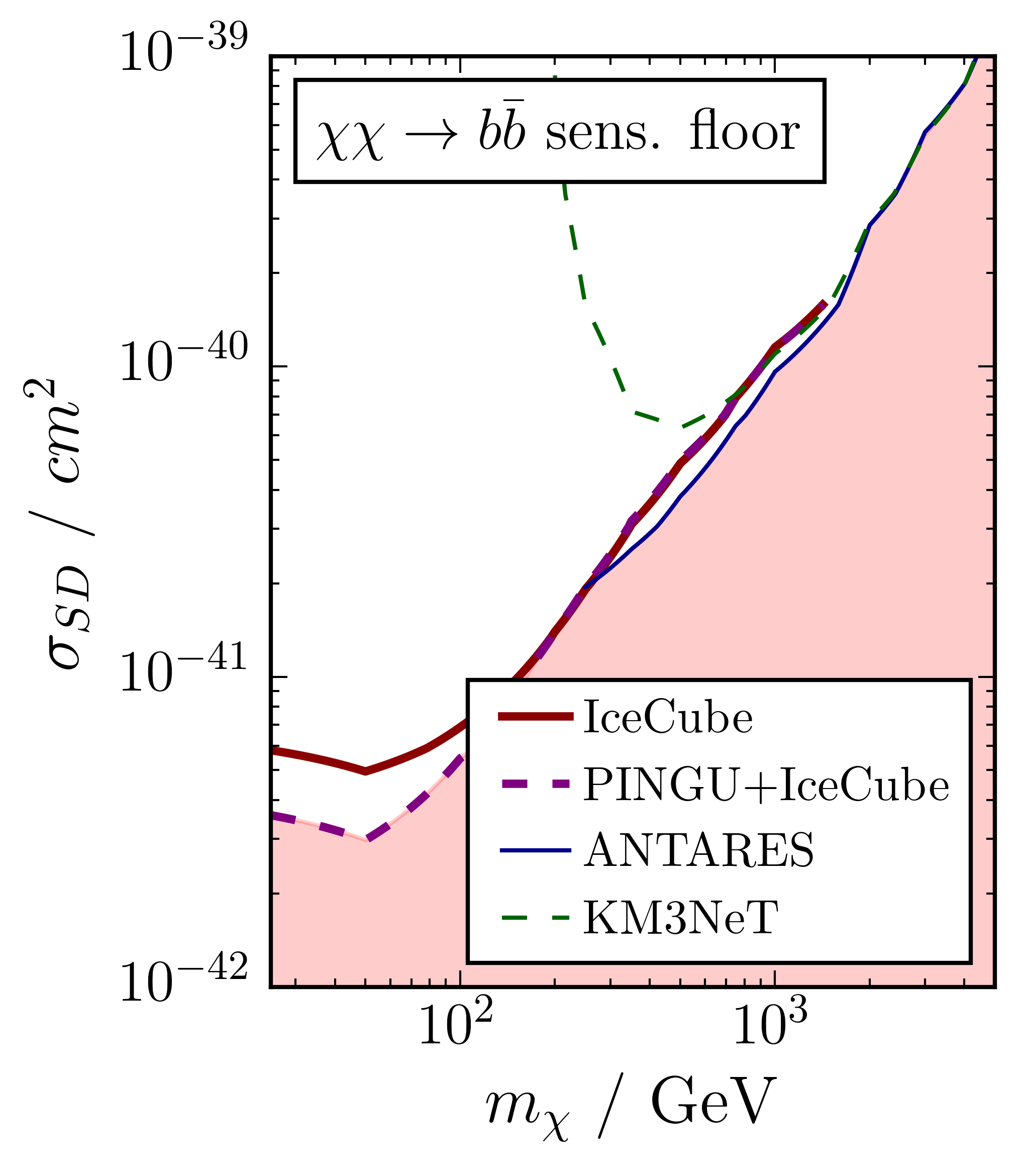}\includegraphics[width=0.32\textwidth]{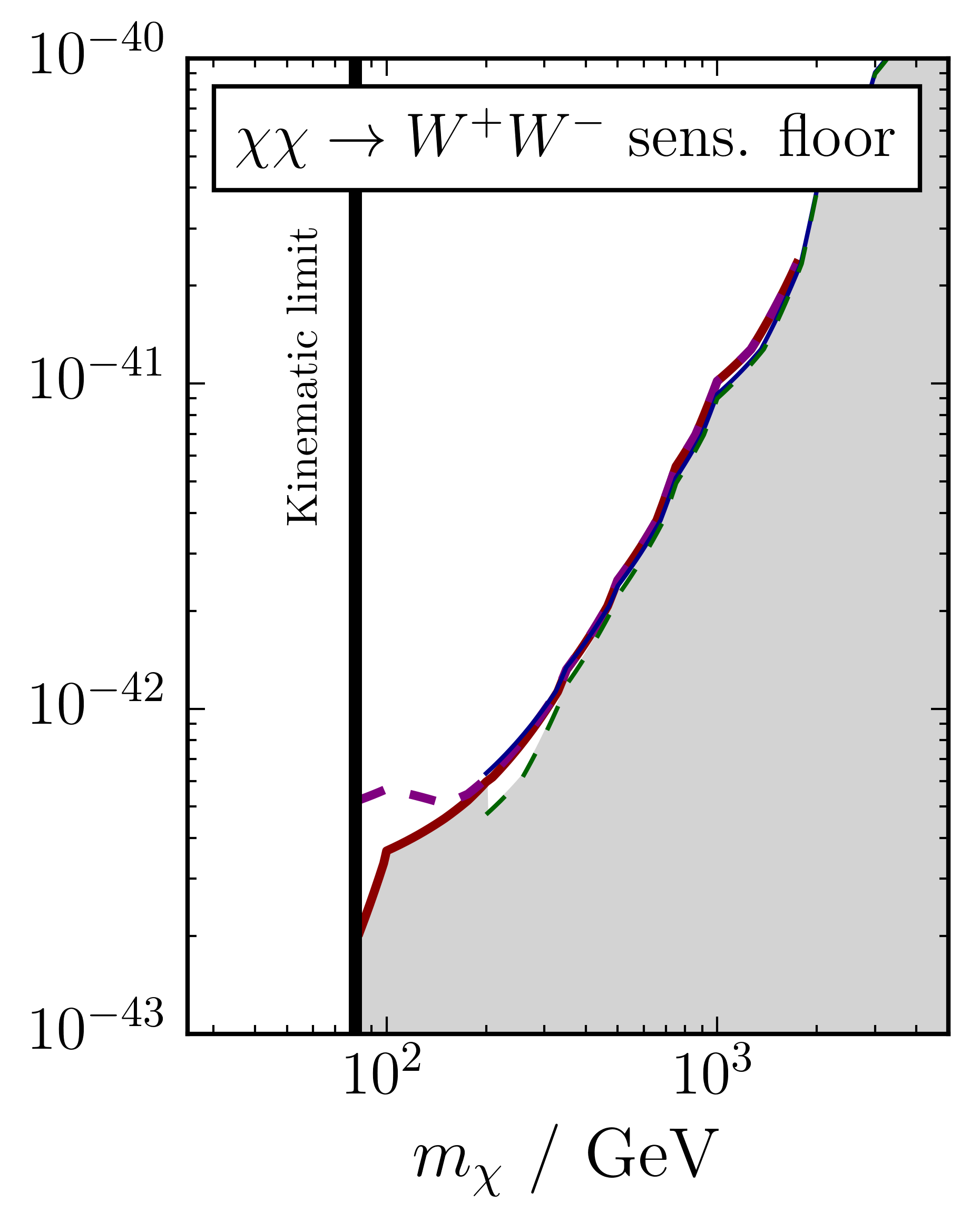}\includegraphics[width=0.32\textwidth]{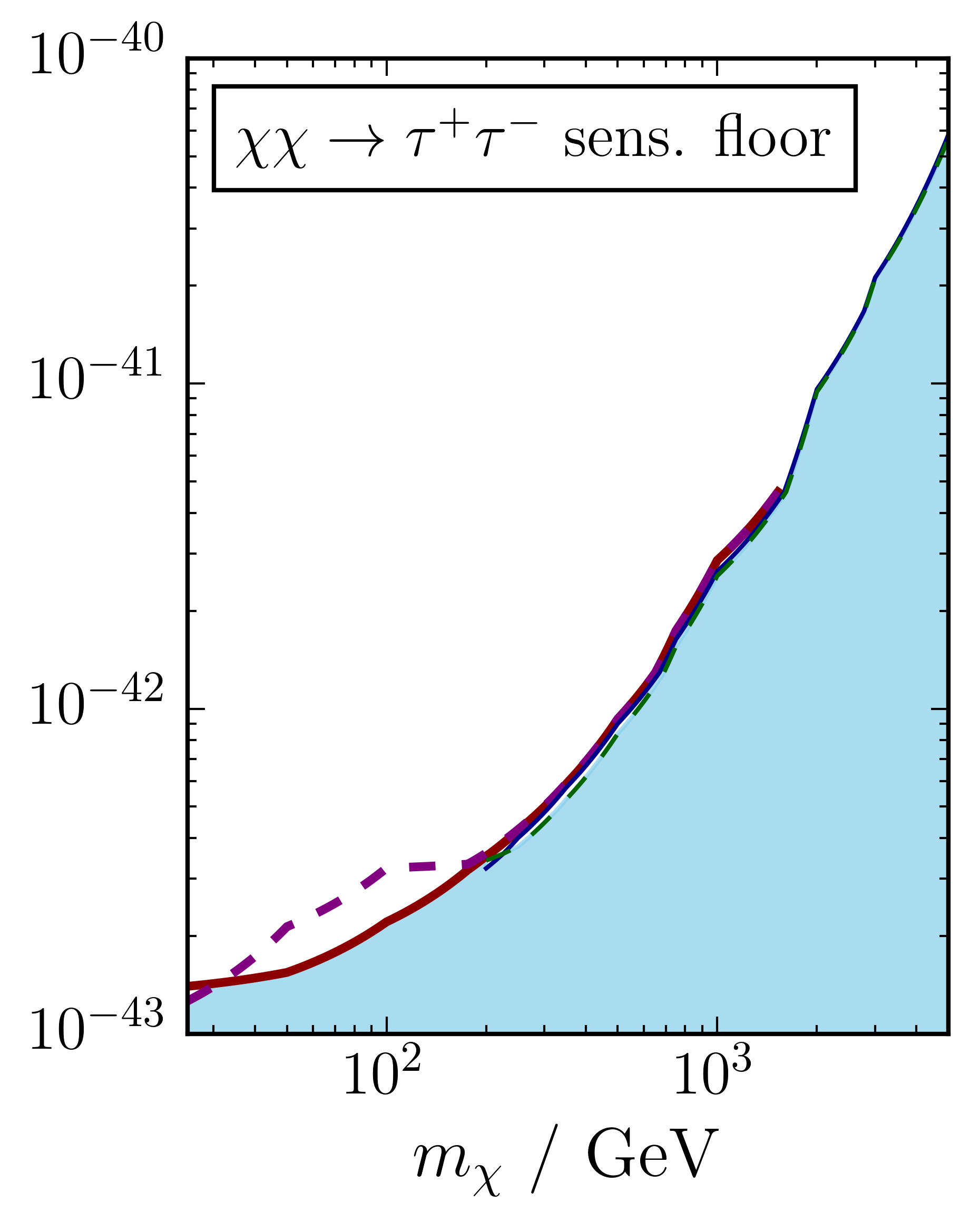}

\par\end{centering}

\caption{The predicted sensitivity floor for each effective area in the $\chi\rightarrow b\bar b$, $\chi\rightarrow W^+W^-$, and $\chi\rightarrow \tau^+\tau^-$  channels. The shaded regions are below the floors for all effective areas. \label{fig:AllFloorsPlot}}
\end{figure}

As a representative dark matter search analysis we consider each WIMP model and define an energy region of interest (ROI) in which to count events.  This ROI is defined such that 90\% of the signal from the WIMP model, after convolution with the relevant effective area, is contained between the upper and lower thresholds of the ROI, with 5\% escaping above and 5\% escaping below.  We then predict the expected signal (S) and background (B) counts in this region. We define the sensitivity floor as being surpassed when S=B.  This analysis method is likely to be slightly weaker than a template likelihood based approach as described in~\cite{Aartsen:2016exj}, though it is stronger than a simple full-sample counting analysis described in~\cite{Aartsen:2016fep}.

\begin{figure}[t]
\begin{centering}
\includegraphics[width=0.6\textwidth]{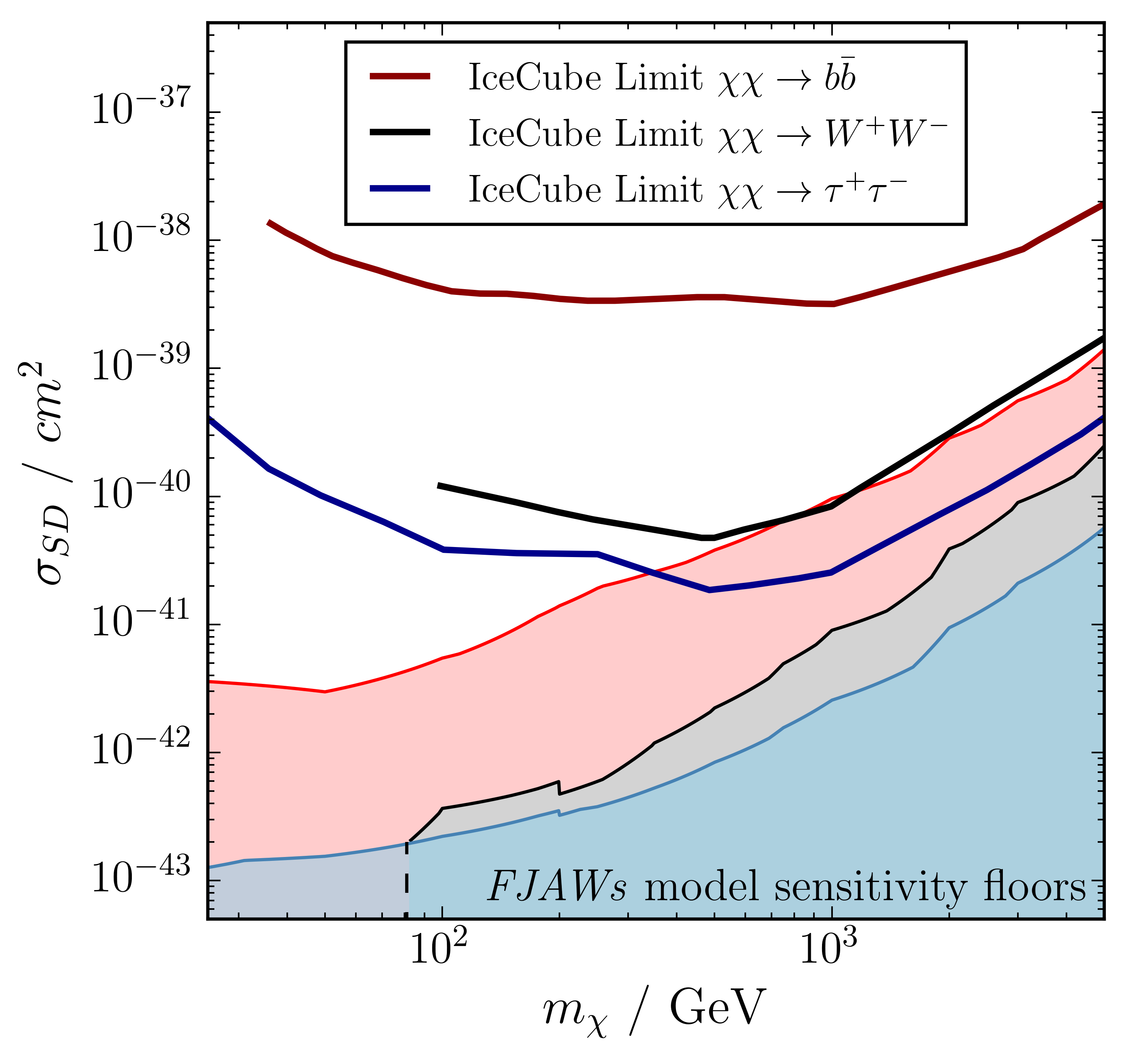}

\par\end{centering}

\caption{The predicted sensitivity floor compared to present experimental limits from ~\cite{Aartsen:2016zhm} in the $\chi\rightarrow b\bar b$ and $\chi\rightarrow W^+W^-$ and $\chi\rightarrow \tau^+\tau^-$  channels.  The lines show the limits, and the shaded regions show parts of parameter space which lie below the sensitivity floor imposed by solar atmospheric neutrinos.\label{fig:NeutrinoFloorPlot}}
\end{figure}

Using the procedure described above, we derive the position of the neutrino floor for each detector and each channel. These are shown in Figure~\ref{fig:AllFloorsPlot}. Since the effective areas differ in their magnitude but not significantly in their shape, the neutrino floor always falls in a similar range in regions where the experiments have sensitivity.   Figure~\ref{fig:NeutrinoFloorPlot} shows the comparison between these sensitivity floors, defined as the minimum accessible spin-dependent cross section given the effective areas considered, and the present leading experimental limits from IceCube from~\cite{Aartsen:2016zhm}.  We see that the solar atmospheric neutrino background will effectively obscure solar dark matter signals less than one order of magnitude below existing spin-dependent cross section limits for some mass points. It is notable that PINGU does not reduce the sensitivity floor beyond IceCube in all channels, because its primary role is to extend the search region to lower energies where the solar atmospheric background is larger. On the other hand, augmenting IceCube with PINGU will slightly increase the detectable rates of both the solar atmospheric neutrinos and of the WIMP signal for some mass points, thus allowing this floor to be reached somewhat sooner. In all cases, our study shows that the sensitivity floor lies approximately one order of magnitude beyond the existing limit in the strongest exclusion regions.

\section{Conclusions\label{sec:Conclusion}}
We have derived the sensitivity floor for indirect solar WIMP annihilation searches, which arises due to interactions of cosmic rays in the atmosphere of the Sun.  Using a new calculation of this flux, called the $\emph{FJAWs}$ model, we predict detectable event rates of around one event per year in existing and proposed neutrino experiments, which introduces a sensitivity floor for indirect dark matter searches using $\chi\rightarrow W^+W^-$, $\chi\rightarrow \tau^+\tau^-$, and  $\chi\rightarrow b\bar{b}$  self-annihilation modes.

Our analysis assumes perfect energy and directional resolution and performs a counting experiment in the energy ROI.  The resolution assumptions are generous, making our statement of the sensitivity floor slightly deep. On the other hand, the use of a full likelihood analysis may lead to a small improvement in sensitivity, though this is unlikely to change our main conclusions or to compensate for the generous assumptions about detector resolution.

We thus conclude that present limits on spin-dependent dark matter interactions from solar WIMP self-annihilation searches are less than one order of magnitude in sensitivity above the floor imposed by the irreducible, high-energy solar neutrino background at some mass points.  Next-generation neutrino telescopes running for multiple years should be able to reach this sensitivity level within their operating lifetimes, thus saturating this particular method of searching for dark matter.

\section*{Note added}
Shortly after the appearance of V1 of this paper on the  arXiv pre-print server, two pre-prints that also propose the neutrino floor were submitted,~\cite{Ng:2017aur} and~\cite{Edsjo:2017kjk}, based on similar work developed independently.  Disagreements in V1 were scrutinized and, through collaboration with the authors of~\cite{Ng:2017aur}, a minor error was identified and corrected in our work, bringing the proposed neutrino floors into closer agreement.

Also during final preparation of this manuscript, an updated IceCube solar WIMP result was published~\cite{Aartsen:2016zhm} which updates the originally cited work~\cite{Aartsen:2016exj} using three years of data from the IC86 detector.  We include here rate estimates and limits based on this new effective area.  However, since the new paper does not provide separated $\nu$ and $\bar\nu$ effective areas, we retain the separate effective areas from~\cite{Aartsen:2016exj} for calculating the neutrino floor.  Since the details of the effective area model have only a minor impact on the location of the floor, this detail does not substantially affect our conclusions.

\acknowledgments

The authors thank David Seckel and Ali Kheirandish for their comments on the text, and Jean DeMerit and Niyousha Davichi for careful proof-reading.  We thank the authors of~\cite{Ng:2017aur} and~\cite{Edsjo:2017kjk} for rewarding scientific discussions. CAD is supported by NSF grants No. PHY-1505858 and PHY-1505855; GdW is supported by the Belgian Inter-University Attraction Pole (IUAP) network `fundamental interactions'; and BJPJ is supported by the University of Texas at Arlington.

\bibliographystyle{unsrt}

\bibliography{biblio}

%\printbibliography[heading=none]

%\begin{thebibliography}{00}
%\bibitem{vernazza}
%Vernazza et al., Structure of the solar chromosphere III.: Models of the EUV brightness components of the quiet Sun (Model C: average quiet Sun) ApJ Supplement Series, vol. 45, 635, 1981.
%\bibitem{christensen}
%Christensen-Dalsgaard, Stellar Structure and Evolution, Lecture Notes, 147, 1993.
%\bibitem{fontenla}
%Fontenla et al., Semiempirical models of the solar atmosphere II. The quiet Sun low chromosphere at moderate resolution. ApJ. 667, 1243, 2007.
%\bibitem{asplund}
%Asplund et al., Line formation in solar granulation,A\&A 417, 751, 2004.
%\bibitem{harvard}
%Gingerich et al., The Harvard-Smithsonian reference atmosphere, Solar Physics, Volume 18, Issue 3, 347, 1971.
%\bibitem{spruit}
%Spruit, A model of the solar convection zone, Sol Phys 34, 277, 1974.
%\bibitem{bs05}
%Bahcall et al., New Solar Opacities, Abundances, Helioseismology, and Neutrino Fluxes, The Astrophysical Journal Letters, 621, L85, 2005.
%\bibitem{geant4}
%Agostinelli et al., Geant4?a simulation toolkit, Nucl. Instrum. Methods Phys. Res. A 506, 250, 2003.
%\end{thebibliography}

\end{document}